\documentclass[twocolumn,english,aps,prb,showpacs,byrevtex,amsmath,amssymb,superscriptaddress]{revtex4-2}

\usepackage{graphicx}
\usepackage{epstopdf}
\usepackage{dcolumn}
\usepackage{bm}
\usepackage{gensymb}
\usepackage{upgreek}
\usepackage{hyperref}
\usepackage{graphicx}
\usepackage{dcolumn}
\usepackage{bm}
\usepackage{gensymb}
\usepackage{upgreek}
\usepackage{booktabs}
\usepackage{hyperref}

\usepackage{amssymb,mathtools}
\usepackage{color}
\usepackage{amsmath} 
\usepackage{xcolor}

\usepackage{tikz,xcolor,hyperref}

\definecolor{lime}{HTML}{A6CE39}
\DeclareRobustCommand{\orcidicon}{%
	\begin{tikzpicture}
	\draw[lime, fill=lime] (0,0)
	circle [radius=0.16]
	node[white] {{\fontfamily{qag}\selectfont \tiny ID}};
	\draw[white, fill=white] (-0.0625,0.095)
	circle [radius=0.007];
	\end{tikzpicture}
	\hspace{-2mm}
}

\foreach \x in {A, ..., Z}{%
	\expandafter\xdef\csname orcid\x\endcsname{\noexpand\href{https://orcid.org/\csname orcidauthor\x\endcsname}{\noexpand\orcidicon}}
}

\begin{document}

\title{Conditions for orbital-selective altermagnetism in Sr$_2$RuO$_4$: tight-binding\\ model, similarities with cuprates, and implications for superconductivity}
\title{Conditions for orbital-selective altermagnetism in Sr$_2$RuO$_4$: tight-binding\\ model, similarities with cuprates, and implications for superconductivity}

\author{Carmine Autieri\orcidA}
\email{autieri@magtop.ifpan.edu.pl}
\affiliation{International Research Centre Magtop, Institute of Physics, Polish Academy of Sciences, Aleja Lotnik\'ow 32/46, 02668 Warsaw, Poland}

\author{Giuseppe Cuono\orcidB}
\affiliation{Consiglio Nazionale delle Ricerche (CNR-SPIN), Unit\'a di Ricerca presso Terzi c/o Universit\'a “G. D’Annunzio”, 66100 Chieti, Italy}

\author{Debmalya Chakraborty\orcidE}
\affiliation{Department of Physics, Birla Institute of Technology and Science - Pilani, K. K. Birla Goa Campus, NH-17B, Zuarinagar, Sancoale, Goa- 403726, India}
\affiliation{Department of Physical Sciences, Indian Institute of Science Education and Research (IISER) Mohali, Sector 81, S.A.S. Nagar, Manauli PO 140306, India}

\author{Paola Gentile\orcidC}
\affiliation{CNR-SPIN, c/o Universit\'a di Salerno, I-84084 Fisciano (Salerno), Italy}

\author{Annica M. Black-Schaffer\orcidD}
\affiliation{Department of Physics and Astronomy, Uppsala University, Box 516, S-751 20 Uppsala, Sweden}

\begin{abstract}
The vibrational modes in Sr$_2$RuO$_4$ easily induce octahedral rotations without tilting. Being on the verge of a magnetic instability, such propensity of octahedral rotation may also produce magnetic fluctuations. In this work, we analyze the long-range magnetic phase diagram incorporating such octahedral rotations and demonstrate the possibility of an altermagnetic phase in Sr$_2$RuO$_4$. 
Using ab-initio calculations, we first study single layer Sr$_2$RuO$_4$ with octahedral rotations, obtaining an orbital-selective $g$-wave altermagnetic phase. We further provide an effective $t_{2g}$ tight-binding model, demonstrating that the $g$-wave altermagnetism is primarily a product of second and third nearest neighbor interorbital hybridizations between the ${\gamma}z$ ($\gamma=x,y$) orbitals, but only a much longer range intraorbital hybridization in the $xy$ orbitals, establishing a strong orbital-selectiveness for the altermagnetism. Notably, by replacing the $xy$ orbital with the $x^2-y^2$ orbital, a similar tight-binding model may be used to investigate the hole-doped cuprate superconductors.
We then study bulk Sr$_2$RuO$_4$, where we find the altermagnetic phase as the magnetic ground state for a range of finite octahedral rotations. In the bulk, interlayer hopping breaks some of the symmetries of the $g$-wave altermagnet, resulting in a $d_{xy}$-wave altermagnet, still with orbital selectiveness.
We also include relativistic effects through spin-orbit coupling and obtain that an effective staggered Dzyaloshinskii-Moriya interaction generates weak ferromagnetism.
Finally, we discuss the implications of the altermagnetic order on the intrinsic superconductivity of Sr$_2$RuO$_4$. Assuming in-plane intraorbital pairing, the altermagnetism favors spin-singlet $d_{x^2-y^2}$-wave or $g$-wave pairing, or their combinations. 
\end{abstract}

\pacs{}
 
\maketitle
 
\section{Introduction}
Altermagnets are recently discovered collinear magnets breaking time-reversal symmetry but still with no net zero magnetization. The altermagnet order is preserved by crystal symmetries, where the spin-up and spin-down sublattices are connected only by rotations (proper or improper and symmorphic or nonsymmorphic). The result is that altermagnets display even-parity wave spin order in reciprocal space, thereby lifting the Kramer's degeneracy in the non-relativistic band structure, leading to unconventional magnetism and also an anomalous Hall effect in symmetry-allowed cases \cite{Smejkal22, Smejkal22beyond, Fakhredine23, Cuono23EuCd2As2, D3NR03681B, D3NR04798A}.

In addition, altermagnets may also host antisymmetric exchange interaction driven by spin-orbit coupling \cite{PhysRevLett.132.176702,PhysRevB.109.024404}. One of the possibilities for antisymmetric exchange interaction is the staggered Dzyaloshinskii-Moriya interaction \cite{autieri2024staggereddzyaloshinskiimoriyainducingweak}, especially in the presence of a single non-magnetic ligand between two magnetic atoms. Overall, the breaking of time-reversal symmetry allows for the presence of multiple band crossings \cite{Fakhredine23} and Weyl points, as already demonstrated in materials such as CrSb \cite{li2024topologicalweylaltermagnetismcrsb}, unstrained CrO \cite{Guo2023} and GdAlSi \cite{nag2023gdalsiantiferromagnetictopologicalweyl}.

Intense activity has also recently been focused on the study of superconductivity in altermagnets. Primarily, proximity-induced superconductivity in superconducting-altermagnet hybrid structures has been studied, revealing exotic effects such as intriguing Josephson effects \cite{Zhang2024,Ouassou23, PhysRevB.108.L060508, sun24, Lu24,fukaya24,mondal2024distinguishingtopologicalmajoranatrivial}, phase-shifted Andreev levels \cite{Beenakker23}, superconducting diode effects \cite{Banerjee24, chakraborty24a}, dissipationless spin-splitting and filtering effects \cite{Giil24,PhysRevB.109.174438}, magnetoelectric effect \cite{Zyuzin24}, topological superconductivity \cite{Li24,Ghorashi24,Zhu23}, and spin-polarized specular Andreev reflections \cite{nagae24, PhysRevB.108.054511}.
Moreover, intrinsic superconductivity has also recently been considered in simple altermagnet models with intriguing consequences, such as finite momentum-pairing at zero applied magnetic field \cite{Rodrigo14, PhysRevB.110.L060508, Shuntaro23, Bose24, sim24, hong24} or field-induced superconductivity \cite{PhysRevB.110.L060508}.
From the point of view of finding materials with both intrinsic superconductivity and altermagnetism, the high-temperature cuprate superconductors\cite{Spalek2023-jg} have so far gathered the most attention, in particular, the parent compound La$_2$CuO$_4$ \cite{Smejkal22,mazin2022notes} and SrRbCuO$_2$Cl$_2$\cite{li2024dwavemagnetismcupratesoxygen}. At high temperatures, La$_2$CuO$_4$ crystallizes in the high symmetry space group I4/mmm (number 139), which does not host altermagnetism, but at low temperatures, a slight in-plane octahedral rotation results instead in the space group 64, which, in combination with spin moments coupling antiferromagnetically, causes breaking of time-reversal symmetry and a natural propensity for altermagnetic order.
Moreover, in the parent compound Nd$_2$CuO$_4$, relevant for electron-doped cuprate superconductivity, both the so-called $T'$ and $T^*$ phases exhibit antiferromagnetism with Kramers degeneracy. However, only a mixed phase, explicitly lacking Kramers degeneracy, has been shown to host superconductivity \cite{PhysRevB.105.014512}.
In addition, signatures of breaking of time-reversal symmetry have been observed in other high-temperature cuprates superconductors \cite{WEBER1990511,doi:10.1126/science.abl8371} and in several other classes of superconductors \cite{Andersen2024}, although this symmetry breaking has often been associated with disorder. 
Finally, hints of coexistence between altermagnetism and superconductivity currently also exist in experimental data on strained RuO$_2$\cite{Ruf2021} and monolayer FeSe \cite{mazin2023induced}.
Overall, these results point to intriguing possibilities for altermagnetism in intrinsic superconductors, but the hunt for a definitive intrinsic altermagnetic superconductor remains ongoing.

\begin{figure}[t!]
\centering
\includegraphics[width=\columnwidth,angle=0]{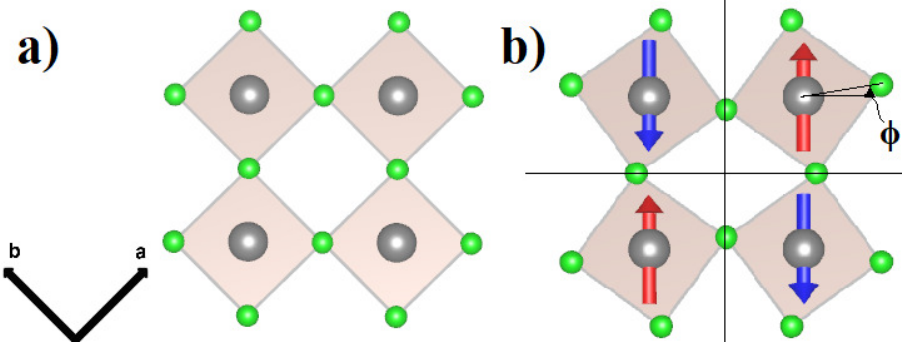}
\caption{(a) Single RuO$_2$ layer of Sr$_2$RuO$_4$ without octrahedral distortion, where native magnetism is absent. (b) Distorted RuO$_2$ layer under the $M$ vibrational mode, presenting a rotation without tilting with angle $\phi$, where magnetism can become present. 
Black lines represent the mirror plane protecting the zero net magnetization in the non-relativistic limit.
Atoms Ru (grey) and O (green) with spin-up (red) and spin-down (blue) electrons on the Ru atoms. Unit vectors \textbf{a} and \textbf{b} represent the lattice vectors of the supercell with 2 in-plane Ru atoms. Octahedral rotations are here enhanced for better visibility.}
\label{crystal_structure}
\end{figure}

In this work, we explore the possibilities and consequences of the intrinsic superconductor Sr$_2$RuO$_4$  being an altermagnet, or at least very close to an altermagnet instability.
Superconductivity has been widely studied in Sr$_2$RuO$_4$ during the last three decades \cite{Maeno1994,Maeno2024,Maeno24a}. The superconducting phase is believed to break time-reversal symmetry \cite{Maeno2024}, and therefore connections to altermagnetism are intriguing. However, the exact symmetry of the superconducting order parameter is still debated, owing to discrepancies between experimental pieces of evidence \cite{Maeno2024,Maeno24a}. For many years, a spin-triplet chiral $p$-wave was the assumed superconducting state \cite{RevModPhys.75.657}, but updated Knight measurements have instead suggested a spin-singlet superconducting state \cite{Pustogow2019}. 
Recent discussions on the superconductivity of Sr$_2$RuO$_4$ have analyzed several open questions concerning in particular the symmetry of the gap function and proposed different options for the symmetry of the superconducting state, in order to reconcile all different experimental findings \cite{Kivelson2020,Valadkhani2024,Maeno2024,Maeno24a,PhysRevB.108.094516,PhysRevResearch.6.043057,Profe2024}. These now include proposals for the superconducting state belonging to the even-parity one-dimensional (1D) irreducible representation $B_{1g}$, resulting in $d_{x^2-y^2}$-wave symmetry \cite{Roising19,Romer19,Romer20}, or the 2D irreducible representation $E_{1g}$ \cite{Maeno2024,Maeno24a} with possible chiral $(d_{xz}+id_{yz})$-wave symmetry, which breaks time reversal symmetry, or nematic symmetry. Additional possibilities are generated due to multiple low-energy orbitals being present \cite{Suh20,Clepkens21,Gingras22} or accidental combinations of different irreducible representations, such as the time-reversal symmetry breaking $d_{x^2-y^2}+ig_{xy(x^2-y^2)}$-wave state \cite{Kivelson2020,Wang22,Yuan23}, combining components from B$_{1g}$ and A$_{2g}$ irreducible representations.
Within this context, an important role is also played by the spin-orbit coupling, which has been demonstrated to induce relevant modifications on Sr$_2$RuO$_4$ Fermi surface, with the induction of a strong momentum dependence, for both orbital and spin character of the low-energy electronic states \cite{Damascelli08}. Furthermore, spin-orbit coupling in Sr$_2$RuO$_4$ leads to the breakdown of pure singlet and triplet states, resulting in a description of the unconventional superconducting state of Sr$_2$RuO$_4$ in terms of spin-orbital entangled eigenstates \cite{Damascelli14}.

It is not only the properties of the superconducting phase of Sr$_2$RuO$_4$ that currently contain many open questions, but its magnetic properties are also not clear. In the early stages after the discovery of superconductivity in Sr$_2$RuO$_4$, it was proposed that ferromagnetic fluctuations would be responsible for both the breaking of time-reversal symmetry and exotic superconductivity \cite{PhysRevLett.79.733}. Instead, incommensurate antiferromagnetic spin-fluctuations were measured for the $\alpha$ and $\beta$ Fermi surfaces \cite{PhysRevLett.83.3320,Raghu_2013} and seen to stay in the doped system \cite{PhysRevB.107.L201107}. The $\alpha$ and $\beta$ Fermi surfaces are dominated by the Ru $d_{xz}$ and $d_{yz}$ (henceforth $\gamma z$, with $\gamma = x,y$) orbitals and are also known to display magnetic spin fluctuations that are highly sensitive to structural changes \cite{PhysRevB.107.144406,PhysRevLett.130.026702}.  
However, antiferromagnetism alone does not break time-reversal symmetry, which is broken in the superconducting phase at least. Altermagnetism, on the other hand, breaks time-reversal symmetry, thus offering an intriguing possibility. The space group I4/mmm (number 139) of Sr$_2$RuO$_4$ does not generate altermagnetic order, due to symmetry reasons, as illustrated in Fig.~\ref{crystal_structure}(a). 
Yet, it is known that the $M$ phononic mode in Sr$_2$RuO$_4$ is soft and just 6~meV above the ground state at 15~K \cite{PhysRevB.57.1236,PhysRevB.76.014505}, inducing lattice displacements in the form of in-plane octahedral distortion without tilting of the RuO$_6$ octahedra. This distortion intriguingly allows for altermagnetism, as shown in Fig.~\ref{crystal_structure}(b).
The antiferromagnetic coupling in ruthenate oxides, and thus also the altermagnetic state, is known to derive from the interplay between octahedral distortions, electronic correlations, and Hund couplings \cite{Sutter2017-rx}.  
We further note that another layer-dependent altermagnetic solution in transition metal oxides has been found in Ca$_2$RuO$_4$, but it has a different crystal space group compared to Sr$_2$RuO$_4$. The magnetic ground state of bulk Ca$_2$RuO$_4$ (space group 61) instead shows orbital-selective altermagnetism with no altermagnetism in the $xy$ orbitals \cite{Cuono23orbital}, while the single layer Ca$_2$RuO$_4$ (space group 7) shows altermagnetism in the $xy$ orbitals \cite{gonzalez2024altermagnetismdimensionalcaruoperovskite}.
On the other hand, Ca$_3$Ru$_2$O$_7$ needs strain to exhibit altermagnetism \cite{leon2025strainenhancedaltermagnetismca3ru2o7}.

In this work, we capitalize on the soft $M$ mode vibrations and consider the properties Sr$_2$RuO$_4$ with small static in-plane octahedral rotations of the RuO$_6$ octahedra. Notably, these in-plane octahedral rotations result in antiferromagnetic fluctuations producing an altermagnetic spin configuration. Since the typical Fermi velocity is higher than a typical sound wave velocity, it is relevant to consider in this way the consequences of a static altermagnetic solution for Sr$_2$RuO$_4$, as we may assume that the electrons feel a static field from the induced magnetic fluctuations due to the lattice distortions. In fact, we are even able to establish that for a range of rotation angles and Coulomb interaction strengths, bulk Sr$_2$RuO$_4$ is an intrinsic altermagnet.
Interestingly, the surface of Sr$_2$RuO$_4$ is actually already known to spontaneously host large such static octahedral distortions, $\phi>7^\circ$ \cite{Damascelli00}. The surface also hosts a magnetic moment along the $c$-axis lower than $0.01~\mu_B$ per Ru atom with an onset temperature larger than 50 K, compatible with the translational symmetry of the crystal and a homogeneous distribution of the magnetism \cite{Fittipaldi2021,Mazzola2024}. This magnetism has so far been theoretically proposed to originate from orbital loop currents \cite{Fittipaldi2021}, but a coinciding study to ours has suggested a possible surface-based altermagnetic mechanism \cite{ramires2025puremixedaltermagnetsintrinsic}. 

In particular, in this work we investigate the static long-range antiferromagnetic phase with ordering $q$-vector equal to ($\frac{\pi}{a_{uc}}$,$\frac{\pi}{a_{uc}}$,0) in the presence of in-plane octahedral rotation, where $a_{uc}$ is the in-plane lattice constant of the Sr$_2$RuO$_4$ unit cell. Using density functional theory (DFT) with added correlation effects, we demonstrate that this leads to a $g$-wave altermagnetic Fermi surface in the single-layer limit of Sr$_2$RuO$_4$, whereas in the bulk, it transitions to a $d_{xy}$-wave altermagnetism due to interlayer hybridization. We further derive an effective $t_{2g}$ tight-binding model for the altermagnet state and demonstrate that the single-layer $g$-wave altermagnetism is due to second and third nearest neighbor (NN) hybridization between the $\gamma z$ Ru orbitals,  while the $xy$ orbitals only experience altermagnetism through seventh NN hopping and is therefore smaller. We note that so far, model Hamiltonian studies have primarily been focused on single-orbital $d$-wave altermagnetism, with intraorbital hopping driving altermagnetism. Instead, here we find $g$-wave altermagnetism in the $t_{2g}$ manifold, with explicitly interorbital hopping driving the altermagnetism. Furthermore, the altermagnetism in Sr$_2$RuO$_4$ is fully orbital selective in the single layer limit, with the $\gamma z$ and $xy$ orbitals experiencing altermagnetic spin-splittings of different origins, and it also retains this orbital selectiveness in the bulk. For the bulk, we also include relativistic effects in terms of spin-orbit coupling and establish the existence of weak ferromagnetism from a staggered Dzyaloshinskii-Moriya interaction. The weak ferromagnetism is orthogonal to the N\'eel vector that we find to be in-plane but would be absent if the N\'eel vector is rotated out-of-plane.

We further discuss the implications on superconductivity from the altermagnetic order. Due to a strong spin-sublattice coupling in altermagnets \cite{Smejkal22beyond,chakraborty2024constraints}, we find that intraorbital in-plane spin-singlet $d_{x^2-y^2}$-wave or $g$-wave pairing symmetries, or time-reversal breaking or preserving combinations thereof, are favored. In contrast, intraorbital $s$-, $d_{xy}$-, and the simplest $p$-wave symmetries are not achievable in fully spin-split altermagnetic Sr$_2$RuO$_4$.
We also note that by replacing the $xy$ orbital with the $x^2-y^2$ orbital, a similar tight-binding model as we derive for Sr$_2$RuO$_4$ may be used to investigate the hole-doped cuprate superconductors.

The rest of this work is organized as follows. In Sec.~II we describe the computational details of the ab-initio calculations.
In Sec.~III we present our results for single layer Sr$_2$RuO$_4$, including deriving a tight-binding model for the resulting $g$-wave altermagnetism, while in Sec.~IV we present the results for bulk Sr$_2$RuO$_4$, including the magnetic phase diagram and the effects of spin-orbit coupling resulting in weak ferromagnetism. In Sec.~IV we discuss the implications of an altermagnetic state on the intrinsic superconductivity of Sr$_2$RuO$_4$. Finally, Sec.~V is devoted to our final remarks and conclusions.

\section{Computational details}
In this work, we perform first-principles calculations of the electronic structure of both single layer and bulk Sr$_2$RuO$_4$ based on density functional theory as implemented in the Vienna \emph{ab-initio} simulation package (VASP) \cite{Kresse93,Kresse96b,Kresse99}.
The pseudopotentials are described using the projector augmented wave method, and the exchange-correlation functional is treated within the local density approximation since it has been found to be more accurate in describing magnetism in ruthenates \cite{Autieri_2016}.
A cutoff energy of 430~eV is applied for the plane-wave expansion and the total energy is converged to  $10^{-5}$ eV/atom. 
We further employ the Liechtenstein approach \cite{Liechtenstein95density} to capture correlation effects, represented by the Hubbard $U$ and Hund's coupling $J_H$ on the $4d$ Ru orbitals. 
In Section III, we use $U=1.0$~eV to emphasize the magnetic properties for the single layer, while in Section IV, for the bulk we scan the value of $U$ from 0.4 to 0.7~eV, keeping $J_H = 0.15U$ as is typically used for the $4d$ ruthenates \cite{Autieri_2016,Roy2015}. Spin-orbit coupling (SOC) is further added within the ab-initio DFT scheme, and thus contains both the on-site and the $k$-dependent parts.

In terms of the crystal structure, we fix the lattice constants to the experimental low-temperature crystal structure of Sr$_2$RuO$_4$ \cite{PhysRevB.57.5067}. Defining $a_{uc}$ and $c_{uc}$ as the lattice constants of bulk Sr$_2$RuO$_4$ using a conventional unit cell, we create a supercell of size $\sqrt{2}a_{uc} \times \sqrt{2}a_{uc} \times c_{uc}$, thus containing 4 Ru atoms for the bulk and 2 Ru atoms for the single layer in order capture the in-plane octahedral distortion.  To capture possible magnetic phases of bulk Sr$_2$RuO$_4$, we scan the in-plane octahedral rotation angle $\phi$ between 0 and 3 degrees, keeping the angle fixed during the calculation of the electronic structure. However, for the Fermi surface calculations, we use a large value $\phi=\arctan{(\frac{2}{25})}\approx 8^{\circ}$ to better illustrate in the figures the symmetries of altermagnetism. To extract the Fermi surfaces, we utilize the Wannier90 software, which can transform Bloch states into Wannier states and also perform maximal localization of the Wannier states \cite{Mostofi:2008_CPC,w90}. We note that, while maximally localization is usually not necessary to fit tight-binding parameters in the Wannier basis, for the altermagnet we find that maximally localization is crucial to obtain an accurate fitting of the band structure.
We sample the irreducible Brillouin zone of the supercell using a $10 \times 10 \times 4$ $k$-point mesh centered on $\Gamma$ for the bulk calculations, while for the single layer case, we use a  $10 \times 10 \times 1$ $k$-point mesh.

Finally, in order to properly describe the altermagnet phase, we switch off the symmetries in the $k$-space and we also constrain the system to have zero net magnetization to avoid spurious numerical errors in the magnetization in all our non-relativistic calculations. The nearby antiferromagnetic phase is similarly captured, while for the ferromagnetic phase, we allow for a finite magnetization.
When we finally incorporate relativistic effects, a net magnetization is again allowed, also for the altermagnetic solution.

\section{Altermagnetism in single layer S\MakeLowercase{r}$_2$R\MakeLowercase{u}O$_4$}\label{sec:3}

We start by investigating the consequences of octahedral rotations in the single layer limit of Sr$_2$RuO$_4$. In particular, we show that such rotations give rise to a $g$-wave altermagnetic Fermi surface. We then provide a minimal tight-binding model to describe this $g$-wave altermagnetism, which turns out to be generated by interband second and third NN hopping parameters in the ${\gamma}$z subsector, and also a longer range intraband hopping in the $xy$ subsector, thereby establishing Sr$_2$RuO$_4$ as a promising candidate for orbital selective altermagnetism.

\subsection{$g$-wave altermagnet Fermi surfaces}

We consider a single layer of Sr$_2$RuO$_4$ composed of the basal plane RuO$_2$, which is sandwiched between two SrO layers. This native single layer of Sr$_2$RuO$_4$ does not host altermagnetism due to crystalline symmetries.
However, if we include an in-plane octahedral rotation of the oxygen cages, the single layer of Sr$_2$RuO$_4$ contains an octahedron that is rotated clockwise and an octahedron that is rotated anticlockwise, see Fig.~\ref{crystal_structure}(b), which may open for altermagnetism. This distortion is favored by the soft $M$ phononic mode and represents a relevant structural instability of Sr$_2$RuO$_4$. With this in-plane octahedral rotation, the space group changes to tetragonal P4/mbm (number 127). A zero net magnetization is still protected by mirror symmetries, with one of the mirror planes determined by the lattice vectors \textbf{c} and \textbf{a}+\textbf{b}, as illustrated in Fig.~\ref{crystal_structure}(b).

\begin{figure}[htb]
\centering
\includegraphics[width=1.13\columnwidth,angle=0]{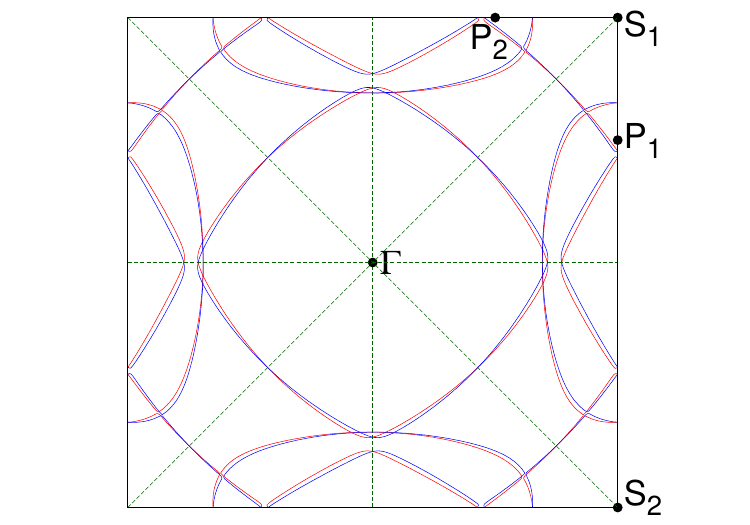}
\caption{Fermi surface of single layer Sr$_2$RuO$_4$ in the $k_x-k_y$ plane for $U = 1.0$~eV and $\phi=8^{\circ}$ with spin-up bands (blue) and spin-down bands (red) in the non-relativistic limit with no SOC. Fermi surfaces show $g$-wave altermagnetism, with nodal lines along $k_x=0$, $k_y=0$, and $k_x=\pm k_y$ indicated by dashed green lines.
We use relatively large $U$ and $\phi$ to increase the magnetic moment to 0.36$\mu_B$ for visualization purposes.
} \label{FS_singlelayer10}
\end{figure}

We calculate the non-relativistic ground state of single layer Sr$_2$RuO$_4$ with octahedral rotation and report the resulting Fermi surface in Fig.~\ref{FS_singlelayer10}. We perform these calculations for relatively large $U$ and octahedral rotation angle $\phi$ in order to increase the non-relativistic spin-splitting for visualization purposes, as the effect of the altermagnetism is less visible on the Fermi surface plots than overall in the band structure.
Studying the Fermi surface in Fig.~\ref{FS_singlelayer10}, we first note that the supercell construction needed to capture the octahedral distortion results in both Brillouin zone folding and a $45^\circ$ rotation compared to the undistorted Sr$_2$RuO$_4$ Brillouin zone. Together with the octahedral distortion and its induced magnetism, this creates several differences between this Fermi surface and the experimental Fermi surface of undistorted bulk Sr$_2$RuO$_4$ \cite{Damascelli00,Bergemann03, PhysRevLett.116.106402,Tamai19, Maeno24a}. 
The Fermi surface of undistorted bulk (and single layer) Sr$_2$RuO$_4$ contain the $\alpha$ and $\beta$ surfaces, originating from hybridized Ru $4d$ ${\gamma}z$ orbitals, and the $\gamma$ surface, originating from the Ru $4d$ $xy$ orbital. For comprehensiveness, we reproduce this Fermi surface in Appendix \ref{app:FS} in the unfolded Brillouin zone, which also establishes that a small in-plane rotation angle mostly preserves the shape of all three Fermi surfaces. However, adding magnetism also somewhat modifies the $\gamma$ Fermi surface. The final result, including finite rotation octahedral angle, magnetism, and the Brillouin zone folding, is in Fig.~\ref{FS_singlelayer10}. Here, all three Fermi surfaces are present, with the $\alpha$ surface centered around $\Gamma$, while the $\beta$ and $\gamma$ surfaces are centered around the Brillouin zone corners, with the latter slightly crossing the $\alpha$ surface, but with no hybridization. The $\gamma$ surface also takes on a more elliptical shape, while the two others have a more square shape.

Regarding the magnetic symmetries, the Fermi surface shows no spin-splitting along $k_x=0$, $k_y=0$, $k_x=k_y$, and $k_x=-k_y$, and there is also a sign change in the spin-splitting when these lines are crossed. Therefore, we conclude that the magnetic phase of single layer Sr$_2$RuO$_4$ is a P-4 $g$-wave altermagnet, using recent notation \cite{Smejkal22beyond}, where P-4 stands for planar with 4-fold rotational repetition of each spin sector, while $g$-wave represents the number of in-plane nodes.
Here, the $g$-wave order is of the type $xy(x^2-y^2)$, and we note that very few material candidates have so far been proposed with this order, but one exception is KMnF$_3$ \cite{Smejkal22beyond}.
We further note that the spin-down Hamiltonian can be obtained by exchanging $k_x$ and $k_y$ in the spin-up Hamiltonian, as the Fermi surface is antisymmetric with respect to the line $k_x=k_y$.

\subsection{Altermagnet spin-split band structure}
\begin{figure}[t!]
\centering
\includegraphics[width=6.2cm,height=\columnwidth,angle=270]{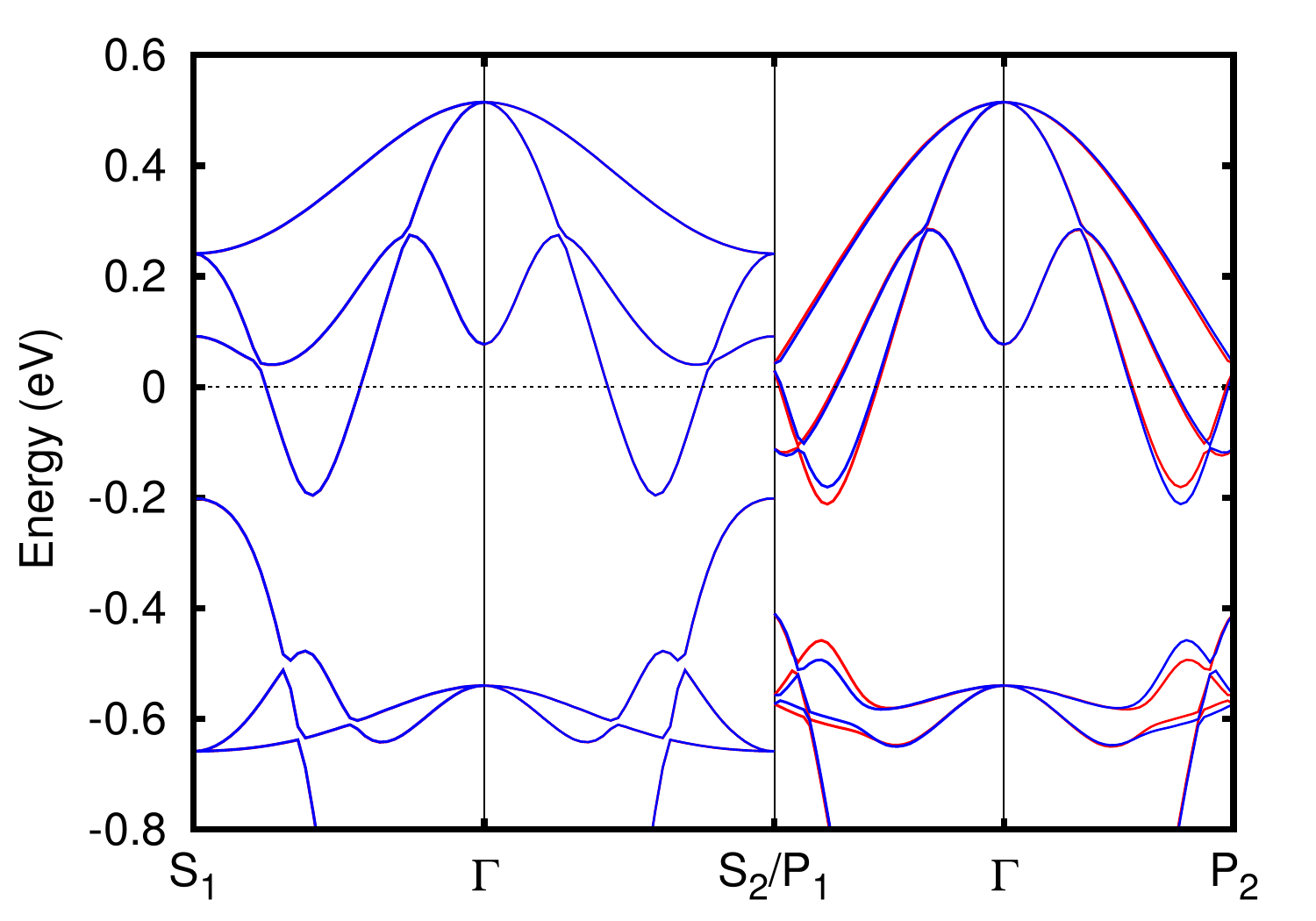}
\caption{Band structure of the single layer Sr$_2$RuO$_4$ for $U=1.0$~eV and $\phi=8^{\circ}$, with spin-up bands (blue) and spin-down bands (red) in the non-relativistic limit with no SOC.}\label{BS_singlelayer}
\end{figure}

To further investigate the altermagnetic phase, we plot in Fig.~\ref{BS_singlelayer} the non-relativistic band structure, both along the altermagnet nodal directions $\Gamma-\rm{S}_i$, with $\rm{S}_1=(0.5,0.5,0)$ and $\rm{S}_2=(0.5,-0.5,0)$ and along the altermagnet anti-nodal directions $\Gamma-\rm{P}_i$ with $\rm{P}_1=(0.50,0.25,0)$ and $\rm{P}_2=(0.25,0.50,0)$, where the spin-splitting is approximately at its largest, as defined in Fig. \ref{FS_singlelayer10}. 
For the altermagnet unit cell with two Ru atoms, we have in total $12$ $t_{2g}$ (spin-resolved) bands from the Ru atoms at low energy, $10$ of these bands are fully included in the plotted energy window between $-0.8$ and $+0.6$~eV, while the other $2$ bands, of $xy$-orbital character, go down up to $-3$~eV also in undistorted Sr$_2$RuO$_4$ \cite{PhysRevB.89.075102, PhysRevB.85.075126}.
In Appendix \ref{app:band} we provide the same band structure plot over a larger energy range, as well as results of the wannierization of the $t_{2g}$ orbitals, needed to extract the Fermi surface already discussed in Fig.~\ref{FS_singlelayer10}. 

We find no spin-splitting along the altermagnet nodal directions $\Gamma-\rm{S}_i$, as expected, but see clear spin-splitting along $\Gamma-\rm{P}_i$. 
Close to the Fermi level, there are 4 (spin-resolved) $\gamma z$-derived bands, making up the $\alpha$ and $\beta$ Fermi surfaces, while there is an additional 2 $\gamma z$-derived bands around $-0.6$~eV. All these show notable spin-splitting. The $xy$-derived bands show some spin splitting around the Fermi level but essentially no splitting for their branch going down to lower energies. As seen in Appendix \ref{app:band}, there is also no notable spin-splitting in the $3z^2-r^2$ bands or any oxygen-derived bands. The $x^2-y^2$ bands have spin-splitting similar to the $xy$ bands.
We thus already here establish the altermagnet phase as strongly orbital-selective\cite{Cuono23orbital}, due to different spin-splitting magnitudes and thus likely different mechanisms driving the non-relativistic altermagnet spin-splitting in different orbitals.
We further note that the ${\gamma}z$ bands are still degenerate in pairs at the $\rm{S}_i$ points, while magnetism splits the eigenvalues of the $xy$ bands at the same points. As a consequence, altermagnetism changes the $xy$ Fermi surface as already established in Fig.~\ref{FS_singlelayer10}.

\subsection{Tight-binding model for $g$-wave altermagnetism}
Having established the existence of an altermagnetic phase in single layer Sr$_2$RuO$_4$, we next extract a tight-binding model to extract where and how altermagnetism enters the low-energy description of the material. We here report the main results, while the full details are reported in Appendix \ref{app:TB}.

A low-energy tight-binding model for Sr$_2$RuO$_4$ is based on the Ru $d$-orbitals in the $t_{2g}$ lower energy subset: $d_{xy}$, $d_{xz}$, and $d_{yz}$. Here we also have two Ru sites, labeled as Ru$_1$ and Ru$_2$. We then define the tight-binding Hamiltonians $H_{11}$ and $H_{22}$ for all processes only involving one Ru atom,  meaning they each contain the on-site energies and the second and third NN hybridization, while the $H_{12}$ Hamiltonian contains NN hybridization, which is a process between the two Ru atoms, see Appendix \ref{app:TB} for a graphical visualization. In Appendix \ref{app:TB} we keep all 12 $t_{2g}$-orbitals, but here we primarily focus on the $\gamma z$ orbitals, as they show most spin-splitting overall in the band structure, while we comment on the altermagnetism in the $xy$-orbitals at the end of this subsection.

Without any magnetism, the $\gamma z$ subsectors of both  Hamiltonians $H_{11}$ and $H_{22}$ can be written as: 
\begin{align} \nonumber
\label{eq:H0}
H^0_{\gamma z} = \quad \quad \quad \quad \quad \quad \quad \quad \quad \quad \quad \quad \quad \quad \quad \quad
\quad \quad \quad \quad \quad \quad \quad
\\  
\left( \begin{smallmatrix}
 \varepsilon_{xz} +2t^{100}_{xz,xz}\cos{k_x} +2t^{010}_{xz,xz}\cos{k_y} &  2t^{100}_{xz,yz}(\cos{k_x}-\cos{k_y}) \\
 2t^{100}_{xz,yz}(\cos{k_x}-\cos{k_y}) & \varepsilon_{xz} +2t^{010}_{xz,xz}\cos{k_x} +2t^{100}_{xz,xz}\cos{k_y}
\end{smallmatrix} \right)
\end{align}
Here $\varepsilon_{xz}$ is the on-site energy of both orbitals $\gamma z$, the terms $t^{100}_{xz,xz}$ and $t^{010}_{xz,xz}$ are intraorbital hybridization terms along the $\bf{a}$ and $\bf{b}$ directions, respectively, while $t^{100}_{xz,yz}$ are interorbital second NN hopping parameters.
We then add the effect of the octahedral distortions. This primary induces a finite term t$^{110}_{xz,yz}$, which is an interorbital third NN hopping arising from the in-plane octahedral rotation. This term is site-dependent because one octahedron is rotating clockwise and another anticlockwise. This hopping has previously also been named $t_{J}$ \cite{Smejkal22} or $t_{am}$ \cite{PhysRevB.110.L060508} in the altermagnet literature. 

Finally, we also need to include a finite spin-splitting, which we do by defining $\Delta_{{\gamma}z}$ as the on-site energy spin-splitting between majority and minority electrons for the $\gamma z$ orbitals. Here we assume, without loss of generality, that spin-up on Ru$_1$ site is the polarization of the majority electrons, while it is spin-down on the Ru$_2$ site.
Taken together, octahedral distortion and finite spin-splitting result in adding to the spin-up $\gamma z$ subsector for the Ru$_1$ atom the Hamiltonian:
\begin{align} 
\label{eq:HAM}
H^{\rm AM}_{\gamma z} = \left( \begin{matrix}
-\frac{\Delta_{{\gamma}z}}{2} &  4t^{110}_{xz,yz}\sin{k_x}\sin{k_y} \\
4t^{110}_{xz,yz}\sin{k_x}\sin{k_y} & -\frac{\Delta_{{\gamma}z}}{2} 
\end{matrix} \right)
\end{align}
To the spin-up $\gamma z$ subsector for the Ru$_2$ atom, instead $-H^{\rm AM}_{\gamma z}$ is added, while in the spin-down subsectors, only the sign of the diagonal $\Delta_{\gamma z}$ term is changed.
Overall, we can write the contributions of Eqs.~\eqref{eq:H0} and \eqref{eq:HAM} to the Hamiltonian for the $\gamma z$  subsector in terms of Pauli matrices for spin, orbital and site as:
\begin{align} 
\label{eq:H0s}
\mathcal{H}^0_{\gamma z}  =& \, \, \, H^0_{\gamma z} \sigma_0^{spin}\sigma_0^{site} \\ 
\nonumber 
\mathcal{H}^{\rm AM}_{\gamma z}  = & -\frac{\Delta_{{\gamma}z}}{2}\sigma_z^{spin}\sigma_0^{orbital}\sigma_z^{site} \\ 
\label{eq:HAMs} & +4t^{110}_{xz,yz}\sin{k_x}\sin{k_y}\sigma_0^{spin}\sigma_x^{orbital}\sigma_z^{site}  \end{align}
We note that altermagnetism is only present if both the $\Delta_{{\gamma}z}$ and t$^{110}_{xz,yz}$ terms are not zero.

Using Eqs.~\eqref{eq:H0s} and \eqref{eq:HAMs}, we define the total Hamiltonian for the low-energy $\gamma z$ subsector as $\mathcal{H}_{\gamma z} = \mathcal{H}^0_{\gamma z} + \mathcal{H}^{\rm AM}_{\gamma z}$. Diagonalizing $\mathcal{H}_{\gamma z}$ results in the eigenvalues of the majority electrons Ru$_1$ and minority electrons Ru$_2$. They differ by a quantity that is the product of the second and third NN interorbital hybridizations, which has the $k$-dependence $(\cos{k_x}-\cos{k_y})$ and $\sin{k_x}\sin{k_y}$, respectively. The product thus has the structure of a $g$-wave orbital: $(x^2-y^2)xy$, which is exactly the altermagnetic order obtained on the Fermi surface of the single layer in Fig.~\ref{FS_singlelayer10}.
We particularly note here that, while most previous altermagnet models in the literature \cite{Smejkal22beyond} have described $d$-wave and $g$-wave altermagnetism only by single-band Hamiltonians with site-dependent (or spin-dependent) intraband hopping parameters, our model Hamiltonian instead generates $g$-wave altermagnetism using a multiorbital Hamiltonian with site-dependent interorbital hopping parameters. Importantly, the material is only altermagnetic if both the spin-slitting $\Delta_{{\gamma}z}$ and third NN interorbital hopping t$^{110}_{xz,yz}$, generated by the octahedral distortion, are not zero. 

This tight-binding model $\mathcal{H}_{\gamma z} = \mathcal{H}^0_{\gamma z} + \mathcal{H}^{\rm AM}_{\gamma z}$ is the minimal model to obtain $g$-wave altermagnetic order in the single layer limit of Sr$_2$RuO$_4$. The full derivation with additional hopping terms and numerical values that are necessary for a realistic description of the full Sr$_2$RuO$_4$ band structure is reported in Appendix \ref{app:TB}. The only free parameters for the ${\gamma}z$ subsector are $\Delta_{{\gamma}z}$ and $t^{110}_{xz,yz}$, where even tiny values are enough to break the time-reversal symmetry of the system and produce the altermagnet order. In Appendix \ref{app:TB} we also establish that the altermagnetic non-relativistic spin-splitting of the $xy$ orbitals is due to intraorbital interactions, but only occurs for seventh NN hopping (and beyond). The altermagnetism in the $xy$ orbitals can therefore be expected to be much smaller than in the $\gamma z$, although the spin-splitting is accidentally still rather large on the Fermi surface, see the $\gamma$ band in Fig.~\ref{FS_singlelayer10}. The free parameters for the $xy$ subsector are $\Delta_{xy}$ and $t^{210}_{xy,xy}$, as shown in Appendix \ref{app:TB}.

To conclude, for single layer Sr$_2$RuO$_4$ we find altermagnetism generated by octahedral distortion, allowing for a new interorbital third NN hopping among the $\gamma z$ orbitals, thus generating altermagnetism in the $\alpha$ and $\beta$ bands. This altermagnetism has $g$-wave symmetry, due to a combination of the distortion-induced interorbital third NN hopping with an intrinsic interorbital second NN hopping term in Sr$_2$RuO$_4$. Altermagnetism in the $xy$ orbital is generally smaller, but still notable on the $\gamma$ Fermi surface, and separately generated by a seventh NN intraorbital hybridization. Thus, we find the $g$-wave altermagnetic phase in single layer Sr$_2$RuO$_4$ to be highly orbital-selective and strongly driven by interorbital processes.

\section{Altermagnetism in bulk S\MakeLowercase{r}$_2$R\MakeLowercase{u}O$_4$}

Having established altermagnetism in single layer Sr$_2$RuO$_4$, we turn to the full bulk structure.
Bulk Sr$_2$RuO$_4$ with in-plane octahedral rotation can be considered through a supercell with two layers and two Ru atoms per layer. Several inequivalent zero net magnetic phases are then possible. Using notation from Ca$_2$RuO$_4$, we name two of these magnetic phases A-centered and B-centered \cite{Porter2022}. In the A-centered phase, octahedra with the same rotations have the same spin and the electronic phase is then an altermagnet. The unit cell of the A-centered magnetic phase can also be obtained with a structure of only 2 Ru atoms.
In contrast, in the B-centered phase, octahedra with the same rotations have the opposite spin and the electronic state is then a Kramers antiferromagnet, since a translation vector between two Ru atoms connects the spin-up and spin-down channel. There exists yet another magnetic phase with ferromagnetic layers coupled antiferromagnetically, which is also a Kramers antiferromagnet, because it also has a translation vector between two Ru atoms that connects spin-up and spin-down channels. As a consequence, we here only consider the A-centered phase as a possible altermagnet solution. 
Symmetry-wise, the space group of bulk Sr$_2$RuO$_4$ changes from space group I4/mmm (number 139) to space group Cmce (number 64) with the in-plane octahedral rotations, which is also the same as the low-temperature phase of the parent high-temperature superconducting compound La$_2$CuO$_4$. 

In this section, we start by establishing the magnetic phase diagram as a function of the octahedral rotation angle, before proceeding with elucidating the properties of the resulting altermagnet phase. In particular, we establish that for a range of small octahedral rotation angles, an altermagnet solution, the A-centered phase, can be the magnetic ground state, and, due to interlayer hybridization, the single layer $g$-wave altermagnetic solution is relaxed to the less symmetric $d_{xy}$-wave structure.

\subsection{Magnetic phase diagram with octahedral rotation}\label{sec:4a}

\begin{figure}[t!]
\centering
\includegraphics[width=5.9cm,height=\columnwidth,angle=270]{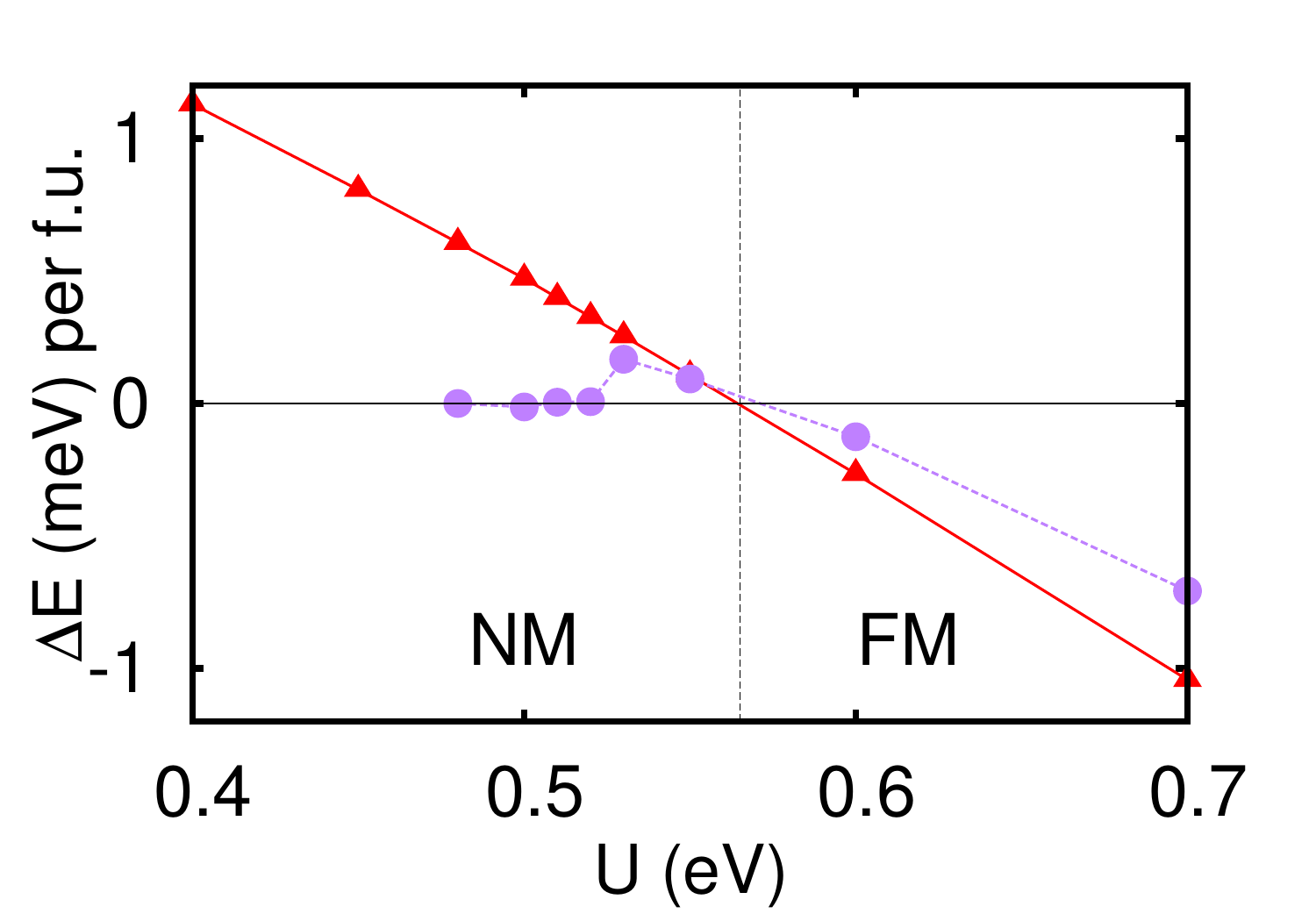}
\caption{Energy difference compared to the nonmagnetic (NM) ground state for a ferromagnetic (FM, red triangles) and conventional antiferromagnetic (AFM, purple circles) states for undistorted bulk Sr$_2$RuO$_4$ as a function of $U$ in the non-relativistic limit without SOC. The ground state is nonmagnetic (NM) up to a critical  $U_{cr}$ where it becomes ferromagnetic (FM). }\label{figure4}
\end{figure}

To establish the possibility of an altermagnetic ground state for bulk Sr$_2$RuO$_4$, we start by reporting the magnetic phase diagram for the undistorted bulk. Previous work \cite{PhysRevResearch.3.033008} has shown that the unstrained bulk transits from a non-magnetic to a ferromagnetic ground state as a function of the Coulomb repulsion $U$. In addition, Sr$_2$RuO$_4$ is close to an antiferromagnetic (AFM) phase, especially for compressive strain \cite{PhysRevResearch.3.033008}.
We reiterate an analysis of the ground state in Fig.~\ref{figure4} by plotting the change in energy for magnetic solutions compared to the nonmagnetic (NM) ground state as a function of the Coulomb repulsion $U$. 
There is a critical value of $U_{cr} = 0.57$~eV, where the ground state changes from NM for $U<U_{cr}$ to ferromagnetic (FM) for $U>U_{cr}$. Moreover, it is possible to establish an antiferromagnetic phase for $U\ge 0.48$~eV, but always at higher energies than the FM phase. Interestingly, the antiferromagnetic phase collapses into the NM solution for lower values of $U$, before forming a plateau in energy for $U=0.48-0.52$~eV, where the AFM phase is essentially degenerated in energy with the NM phase. 

\begin{figure}[htb]
\centering
\includegraphics[width=5.9cm,height=\columnwidth,angle=270]{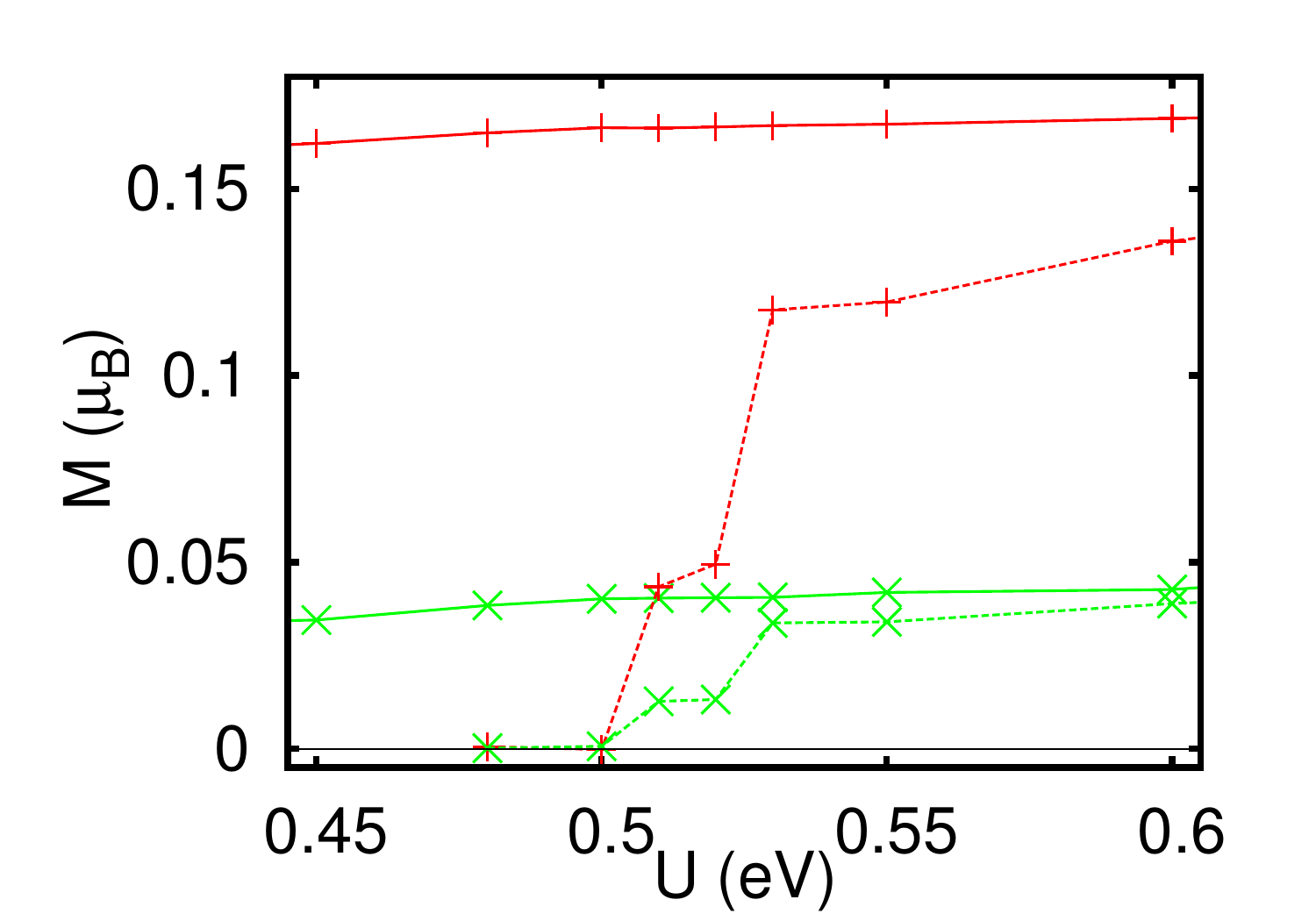}
\caption{Orbital-resolved local magnetization as a function of $U$ for the ferromagnetic (FM, solid lines) and antiferromagnetic (AFM, dashed lines) phases, with the magnetization of the $xy$ orbitals (red) and $\gamma z$ (green) in the non-relativistic limit with no SOC.
}\label{figure5}
\end{figure}
We also report on the orbital-resolved magnetic structure of the FM and AFM phases, as the altermagnet has a strong orbital-selectiveness and hence such structure is likely present in the FM and AFM phases as well. In Fig.~\ref{figure5}, we report the calculated local magnetic moments originating from the different orbitals using the occupation matrix of the Coulomb repulsion $U$. As seen, the local magnetic moments in the FM phase (solid lines) are always larger than in the AFM phase (dashed lines) and also higher on the $xy$ orbitals (red lines). Therefore, the main driver for the FM phase is the magnetization coming from the $xy$ orbitals. However, once we include the in-plane octahedral rotations to stabilize the altermagnetic phase, we obtain magnetization values similar to the  AFM moments in Fig. \ref{figure5}. It is here important to remember that for the ${\gamma}z$ bands the non-relativistic spin splitting depends on the co-presence of both $\Delta_{\gamma z}$ and $t^{110}_{xz,yz}$, while the magnetic moments depend only on $\Delta_{xz}$. The same is valid for the $xy$ bands. Therefore, the non-relativistic spin-splitting and magnetic moments become largely unrelated.

\begin{figure}[htb]
\centering
\includegraphics[width=5.9cm,height=\columnwidth,angle=270]{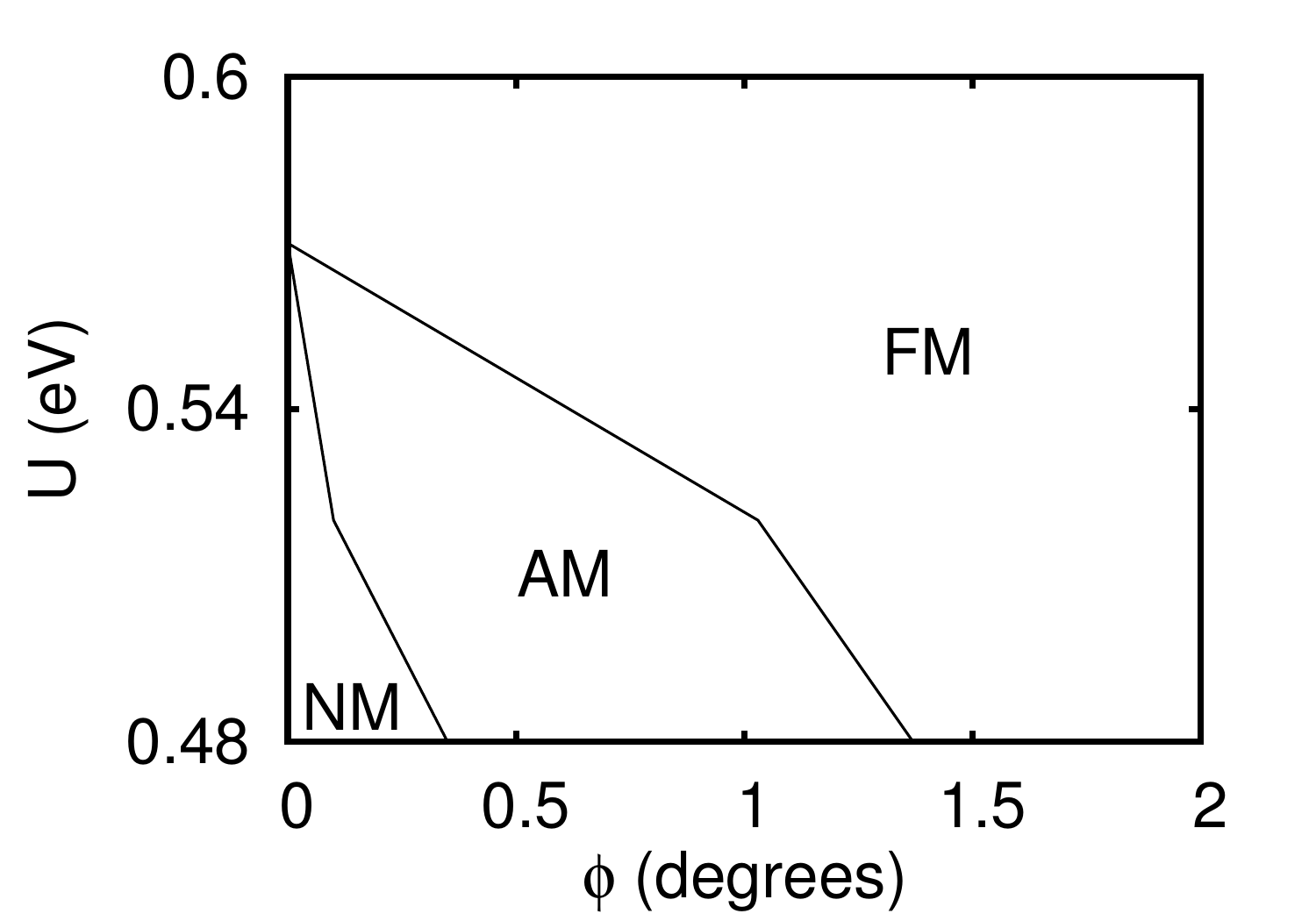}
\caption{Magnetic phase diagram of bulk Sr$_2$RuO$_4$ as a function of Coulomb repulsion $U$ and in-plane octahedral rotation angle $\phi$ in the non-relativistic limit with no SOC. Allowing for small in-plane octahedral rotations, the altermagnetic (AM) phase is stabilized between the non-magnetic (NM) and ferromagnetic phase (FM).}\label{figure6}
\end{figure}

Having established the FM and AFM phases in the undistorted crystal, we turn on in-plane octahedral rotations and map out the phase diagram as a function of both $U$ and the rotation angle $\phi$ in Fig.~\ref{figure6}. An example of the underlying calculated ground state energies as a function of rotation angle $\phi$ for a specific $U$ is reported in Appendix \ref{app:magphase}. 
We obtain that even infinitesimally small rotations can stabilize the altermagnetic phase energetically for a small range of $U$-values around $U=0.57$~eV. In addition, for smaller $U$ we find an increasingly large region of altermagnetism for finite but still small angles. This altermagnetic phase turns into the NM phase for zero angle, while larger angles result in the FM phase. The overall rise of magnetism with increased octahedral rotations can be explained by a bandwidth reduction driven by a diminution of the first NN hopping parameters, which are the primary components of the bandwidth.

To summarize, we find altermagnetism to be possible as the ground state of bulk Sr$_2$RuO$_4$ in the presence of small, or even infinitesimal, octahedral rotations, but where the actual ground state is sensitive to the exact value of the Coulomb repulsion $U$. Intriguingly, the altermagnet phase requires the smallest octahedral rotations around $U_{cr}$ for entering the FM phase, which testifies to a close competition between all magnetic states and that an altermagnetic state is particularly favored by small to moderate octahedral rotations. Furthermore, since the altermagnetism is driven by a coupling between the magnetic and structural properties, we can further expect a strong spin-phonon coupling in this system.

\subsection{Altermagnetic spin-splitting}
To further understand the altermagnetic phase in bulk Sr$_2$RuO$_4$ and how it compares to the single layer, we first note that in Sr$_2$RuO$_4$ we have two subsets of bands: the $\gamma z$ orbitals, where altermagnetism occurs due to interorbital second and third NN processes and therefore is the main source of altermagnetism, and the $xy$ orbitals where altermagnetism also occurs only in an intraorbital seventh NN process.
However, in the bulk, these two subsets of bands are coupled by interlayer hopping parameters, which are of the size $8\tilde{t}^{\frac{1}{2}\frac{1}{2}\frac{1}{2}}_{xy,{\gamma}z}\approx 100$~meV \cite{PhysRevB.74.035115} and by spin-orbit coupling also equal to 100~meV \cite{PhysRevB.109.L241107}, where $\tilde{t}^{\frac{1}{2}\frac{1}{2}\frac{1}{2}}$ indicate the NN interlayer hopping (for more info see notation in Appendix \ref{app:TB} and literature\cite{PhysRevB.74.035115}). 
Beyond these two effects, the $xy$ orbitals stay decoupled from the $\gamma z$ orbitals. We consider the spin-orbit coupling in the next subsection by allowing for relativistic effects, while here we focus on the consequence of interlayer hopping or hybridization.

Considering the interlayer hopping, we first of all note that even an infinitesimal interlayer hopping parameter breaks the spin-degeneracy of the Fermi surface along the nodal lines $k_x={\pm}k_y$ or along the $\Gamma-\rm{S}$ path in the Brillouin zone. This means a finite spin-spin splitting along the $\Gamma -\rm{S}$ paths in the Brillouin zone, and, importantly, that the altermagnet cannot have a full $g$-wave symmetry. The symmetry is instead lowered to a $d_{xy}$-wave altermagnet for bulk Sr$_2$RuO$_4$. The tight-binding model reported above for the $g$-wave altermagnetic single layer can also be transformed into a $d_{xy}$-wave altermagnetic bulk solution just by adding interlayer hopping, see Appendix \ref{app:TB}.
\begin{figure}[htb]
\centering
\includegraphics[width=1.13\columnwidth,height=0.8\columnwidth,angle=0]{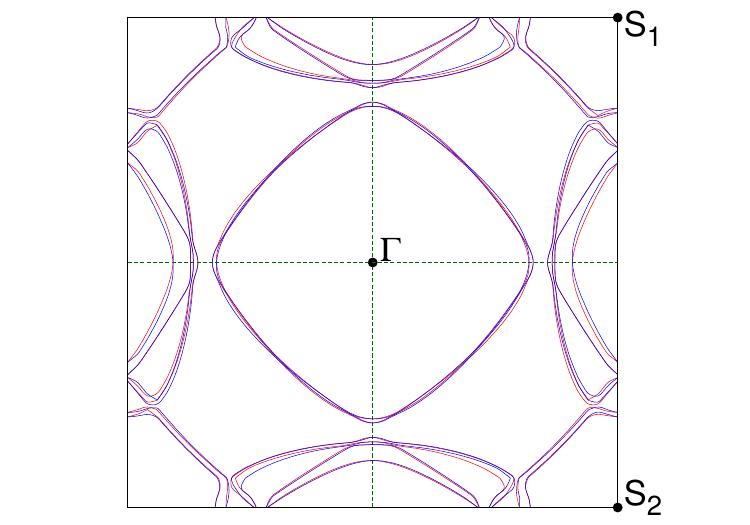}
\caption{Fermi surface of bulk Sr$_2$RuO$_4$ in the $k_z = 0$ plane for $U = 1.0$~eV and $\phi= 8^{\circ}$ with spin-up bands (blue) and spin-down bands (red) in the non-relativistic limit with no SOC. Fermi surfaces show $d_{xy}$-wave altermagnetism with nodal lines along $k_x=0$ and $k_y=0$, indicated by dashed green lines. We use relatively large $U$ and $\phi$ to increase the magnetic moment to 0.36$\mu_B$ for visualization purposes.
}\label{FS_bulk}
\end{figure}

To further study the altermagnet symmetry in bulk Sr$_2$RuO$_4$, we report the Fermi surface at $k_z=0$ in the altermagnet phase in Fig.~\ref{FS_bulk}. A finite spin-splitting is present across the entire Brillouin zone and in all Fermi sheets. 
We further verify in Fig.~\ref{FS_bulk} the overall $d_{xy}$-wave symmetry of the altermagnet spin-splitting, with dashed lines indicating the nodes. We however still have to remember that it originates from a single-layer $g$-wave altermagnetic phase modulated by relatively small interlayer hopping parameters. Therefore, while, the non-relativistic spin-splitting that was originally forbidden in the $g$-wave along the $\Gamma-\rm{S}$ $k$-path in the single layer is now allowed, the spin-splitting along $\Gamma-\rm{S}$ is still relatively small. Hence, the $d_{xy}$-wave symmetry cannot host its maximum spin-splitting along $\Gamma-\rm{S}$ as is usually the case for a $d_{xy}$-wave solution. The overall structure is therefore that of a formal $d_{xy}$-wave with an overlaid $g$-wave modulation originating in the single layer physics. 
Interestingly, with Sr$_2$RuO$_4$ having $g$-wave altermagnetism for a single layer and $d_{xy}$-wave altermagnetism in the bulk, there is a layer-dependent altermagnetism, since the crystal symmetries change as a function of the number of Sr$_2$RuO$_4$ layers. We provide a complementary band structure plot for the altermagnet phase in bulk Sr$_2$RuO$_4$ in Appendix \ref{app:bulkband}. There we also extract the altermagnetic spin splitting at low energies, including at the Fermi level. For example, for $U=1.0$~eV and $\phi=8^{\circ}$ we find a non-relativistic spin-splitting of the ${\gamma}z$ bands as high as $7$~meV at the Fermi level. These values are thus the approximate spin-splitting present on the $\alpha$ and $\beta$ bands. While these are larger octahedral distortions than expected and also larger $U$ values than where the altermagnet phase is naturally the ground state, it illustrates that altermagnetism can create a substantial spin-splitting in bulk Sr$_2$RuO$_4$, especially when compared to the small energy scales relevant for superconductivity, estimated to have an energy gap of just $350~\mu$V \cite{doi:10.1073/pnas.1916463117}.

\subsection{Weak ferromagnetism}
As a final stage in studying the magnetic properties of bulk Sr$_2$RuO$_4$, we include SOC and thereby relativistic effects on top of the $d_{xy}$-wave altermagnetic bulk solution. We find that the SOC slightly increases the spin-splitting properties, but it has little impact on the overall orbital selectivity. Its main effect is instead on the rotation of the spin orientation. 
We find that the interplay between spin-orbit coupling and altermagnetism assumes the form of a staggered Dzyaloshinskii-Moriya interaction (DMI) for the coupling of in-plane atoms. The interplay between different layers is much weaker and we therefore neglect this effect in our discussions.
Staggered DMI is known to be one of the mechanisms responsible for the weak ferromagnetism in altermagnets\cite{autieri2024staggereddzyaloshinskiimoriyainducingweak}. Other forms of spin-orbit effects, antisymmetric with respect to the exchange of the spins, are also possible; however, when there is a single ligand atom connecting the magnetic atoms, the SOC tends to have the staggered DMI form \cite{autieri2024staggereddzyaloshinskiimoriyainducingweak}. A more general form of SOC in altermagnets has also recently been investigated \cite{PhysRevLett.132.176702,PhysRevB.109.024404}.

\begin{figure}[t!]
\centering
\includegraphics[width=\columnwidth,angle=0]{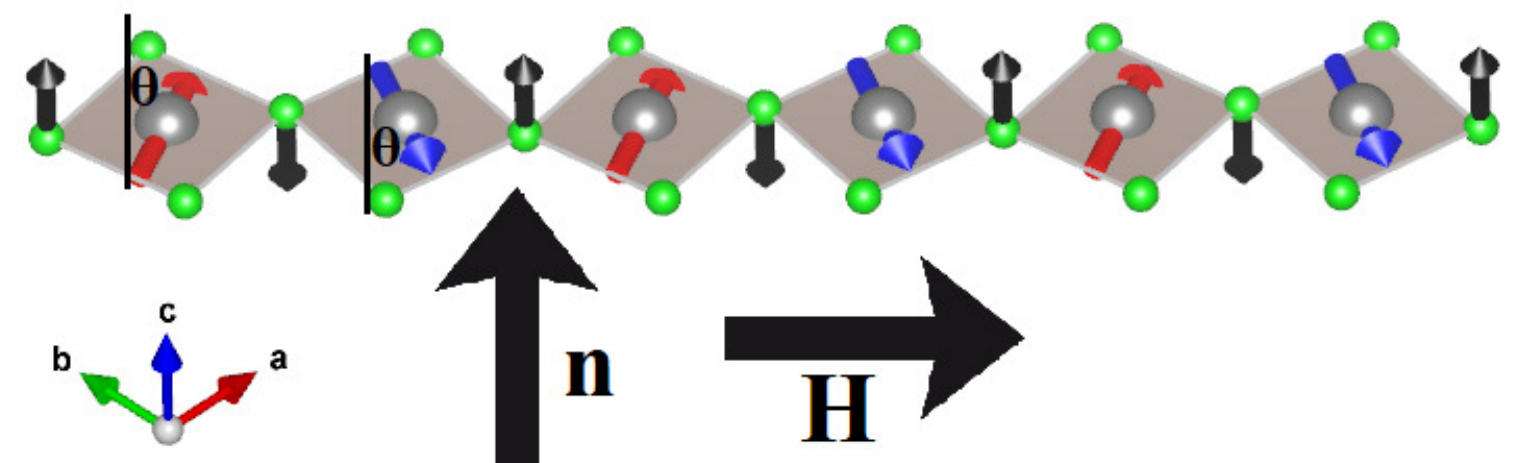}
\caption{Schematic illustration of the in-plane octahedral rotation in a RuO$_2$ layer generating weak ferromagnetism in the presence of spin-orbit coupling. Spin-up (red) and spin-down (blue) moments on the Ru atoms (grey) lie in the $ab$ plane, but tilted out-of-plane by a small angle $\theta$. Staggered DMI vectors (black) are illustrated on the planar oxygen atoms (green), while other DMI vectors on the other oxygen atoms are not shown. Indicated is also the N\'eel vector \textbf{n}, oriented in the $ab$ plane, and the Hall vector \textbf{H} orthogonal to it.
Here, the vectors \textbf{a}, \textbf{b} and \textbf{c} represent the lattice vectors of the supercell with 2 Ru atoms.}\label{weakFM}
\end{figure}

Once we include SOC, we find in our calculations that the N\'eel vector lies in the $ab$-plane in the magnetic ground state. The energy difference between in-plane and out-of-plane N\'eel vectors is $0.2$~meV per formula unit. As we neglect the interlayer spin-orbital effects and only consider the intralayer DMI, the two layers in the unit cell of the bulk have the same properties. 
We schematically illustrate the resulting magnetic order in Fig.~\ref{weakFM}. As seen, the altermagnetic moments on the Ru atoms lie in the $ab$-plane, but with a small out-of-plane tilt $\theta$. This canting of the altermagnetic moments is caused by an out-of-plane DMI vector on the oxygen atoms and results in overall weak ferromagnetism. If, on the other hand, the N\'eel vector is along the $z$-axis, there is no weak ferromagnetism from the staggered DMI.

To further analyze the effect of the resulting DMI interaction, we define $J$ as the antiferromagnetic exchange coupling between the spins with module $S$ generated by the altermagnetism and $D_z$ as the only nonzero component of the DMI. We here note that $D_z$ would be zero without the in-plane octahedral rotation due to the Moriya rules \cite{PhysRev.120.91}. 
With the N\'eel vector \textbf{n} in the $ab$-plane, the total magnetic energy can be described as:
\begin{equation}  \label{total_energy} 
E(\theta)=-4JS^2\cos (2\theta)+4D_zS^2 \sin (2\theta)\quad {\rm if} \quad {\bf n} \perp \hat{{\bf z}},
\end{equation}
where $\theta$ is the canting angle of the spins with module $S$. 
We can also calculate the angle $\overline{\theta}$ that minimizes the total energy by deriving Eq.~(\ref{total_energy}). Considering that $\frac{D_z}{J} \ll $1, we find from this the magnetization to be:
\begin{equation}\label{MZ}
M_{xy}=2S\sin(\overline{\theta}) \approx 2S\overline{\theta} \approx -\frac{SD_z}{J}; \quad M_{z} =0
\end{equation}
Here $\overline{\theta}$ becomes the canting angle reported in Fig.~\ref{weakFM}. 
Since the A-centered magnetic phase in Sr$_2$RuO$_4$ can be obtained with just two magnetic atoms in one layer periodically repeated, the weak ferromagnetic moment of bulk Sr$_2$RuO$_4$ is just the sum over two layers.
We further note that in the field of altermagnetism, the direction of the magnetization is also defined as the Hall vector \textbf{H}, which is always orthogonal to the N\'eel vector in systems involving staggered DMI with two atoms in the unit cell \cite{autieri2024staggereddzyaloshinskiimoriyainducingweak}. 

Given the relatively large SOC in Sr$_2$RuO$_4$, we find that the canting angle can reach up to 3$^{\circ}$ in the weak ferromagnetic phase. Despite this large canting, with respect to other weak ferromagnets \cite{autieri2024staggereddzyaloshinskiimoriyainducingweak}, the overall magnetization is still small due to the small magnetic moments.
Further, the overall gain in the total energy for the spin canting can be calculated as:
\begin{equation}   
E(0)-E(\overline{\theta}) \approx -2JS^2(2\overline{\theta})^2-4D_zS^2(2\overline{\theta})=2S^2\frac{D_z^2}{J}
\end{equation}
which is extremely small, since  $\frac{D_z}{J}<<$1.
The energy gain is only of the order of ${\mu}$eV in materials with 3$d$ electrons, while it can be of the order of $0.1$~meV in 4$d$ electron systems. Therefore, extremely accurate calculations are required to detect this spin canting.
Analogously to the intraplane DMI analyzed above, another SOC-mediated antisymmetric spin-spin interaction is also present interplane, but with a weaker magnitude. Therefore, it will only produce an even weaker ferromagnetism along the z-axis.

\section{Implications of altermagnetism on superconductivity}

Having established the possibility of altermagnetism in both single layer and bulk Sr$_2$RuO$_4$, we finally discuss its implications on the superconducting phase of Sr$_2$RuO$_4$. We focus on the implications on the bulk superconducting phase, since the single layer of Sr$_2$RuO$_4$ has not yet been experimentally exfoliated.
As discussed in Sec.~\ref{sec:4a}, the spin-splitting of the energy bands near the Fermi level, relevant for superconductivity, strongly depends on the angle $\phi$ and for $\phi=8^{\circ}$ it is of the order of 10~meV in bulk Sr$_2$RuO$_4$. This is between 1 and 2 orders of magnitude larger than the superconducting gap in Sr$_2$RuO$_4$, which has been experimentally estimated to be approximately $350~\mu$V \cite{doi:10.1073/pnas.1916463117}. Thus, even if the angle $\phi$ is substantially smaller in Sr$_2$RuO$_4$, it is reasonable to consider that the altermagnetic spin structure may strongly influence the superconducting state. In fact, as discussed in Appendix E, the spin-splitting for angles $\phi=0.5-1^{\circ}$ should be of the order of 1 meV, which is still larger than $350~\mu$V.

The most direct effect of altermagnetism on superconductivity is that it puts constraints on the possible symmetries of any intraorbital superconducting pairing in Sr$_2$RuO$_4$ \cite{chakraborty2024constraints}, due to the distinct spin-sublattice coupling of altermagnets \cite{Smejkal22beyond}. This is because, in the altermagnetic phase, the two Ru atoms in the extended unit cell are essentially fully spin-polarized on the energy scale relevant to superconductivity. For the single layer, these two Ru atoms, Ru$_{1}$ and Ru$_{2}$, thus hold opposite spin electrons on NN sites. In the bulk, interlayer coupling introduces a hybridization between these NN sites, but this hybridization is still small compared to the in-plane spin-polarization. As a consequence, even in the bulk, the low-energy electrons on NN Ru$_{1}$ and Ru$_{2}$ atoms have predominantly different spins. 
In order to have spin-singlet superconducting pairing, opposite spins necessarily need to pair. If we then only consider intraorbital and in-plane pairing, onsite $s$-wave pairing is thus not allowed in the altermagnetic configuration. Instead, $d_{xy}$-wave spin-singlet pairing between the in-plane Ru$_{1}$ and Ru$_{2}$ sites is likely, as its spin-configuration is favored by the altermagnetism \cite{chakraborty2024constraints}. In contrast, the other $d$-wave spin-singlet symmetry, $d_{x^2-y^2}$-wave, will need pairing between Ru$_{1}$ and Ru$_{1}$, which instead share the same spin polarizations, and hence is again not be allowed.

The symmetries outlined above are given relative to the extended $\sqrt{2}a\times\sqrt{2}a \times c$ unit cell of Sr$_2$RuO$_4$, as shown in Fig.~\ref{figure9}. The conventional (smaller) unit cell is rotated by $45^\circ$, and thus, by altermagnetism allowed, spin-singlet intraorbital $d_{xy}$-wave symmetry corresponds to a $d_{x^2-y^2}$-wave state. Thus, the altermagnetic phase of Sr$_2$RuO$_4$ is likely to host $d_{x^2-y^2}$-wave superconductivity, if we consider only the shortest range spin-singlet pairing possibilities. However, longer-range pairings are also a possibility. In particular, $g$-wave spin-singlet intraorbital pairing involves pairing between opposite spins of Ru$_{1}$ and Ru$_{2}$ located between fourth NNs and is hence also favored by the altermagnet phase. 
If we additionally allow for spin-triplet pairing, the same spin-sublattice coupling of the altermagnet prohibits the shortest range spin-triplet $p$-wave symmetry. Taken together, due to the spin-sublattice coupling of the altermagnet, the most likely possible superconducting pairing in the altermagnetic phase of Sr$_2$RuO$_4$ are spin-singlet $d_{x^2-y^2}$-wave pairing, $g$-wave, or combinations thereof such as the time reversal symmetry preserving and generally nematic $d_{x^2-y^2}+g$-wave state or the time-reversal symmetry breaking (but not chiral) $d_{x^2-y^2}+ig$-wave state. These results agree well with recent proposals based on experimental measurements \cite{Maeno2024}. We further note that this analysis involves only intraorbital and in-plane pairing involving the $\gamma z$ orbitals that are strongly altermagnetic. The constraints on the superconducting pairing will not be as strong if the intraorbital pairing occurs instead primarily in the $xy$-orbital where altermagnetism is weaker. In addition, interorbital superconducting pairings are also currently discussed \cite{Maeno24a,Suh20,Clepkens21,Gingras22}. We leave the discussion on the possible influence on altermagnetism on interorbital, and possibly also out-of-plane, superconducting pairing as a future outlook.

Beyond constraining the superconducting symmetries due to spin-sublattice locking, our findings of possible altermagnetism in Sr$_2$RuO$_4$ open up additional future directions in the study of superconductivity in Sr$_2$RuO$_4$ and related materials. 
One important direction concerns one of the most salient features of superconducting Sr$_2$RuO$_4$, which is its broken time-reversal symmetry. Usually, time-reversal symmetry breaking is taken as a signature of a multicomponent superconducting order parameter. However, our results open up the new possibility that the time-reversal symmetry is instead broken due to the presence of altermagnetism and might thus not be tied to superconductivity. In fact, recent experimental data indicate that time-reversal symmetry breaking persists well above the superconducting transition temperature \cite{Fittipaldi2021,Mazzola2024}, which may be indicative of the importance of a magnetic state in Sr$_2$RuO$_4$. Altermagnetism is then a prime candidate due to its zero net magnetization or only very weak ferromagnetism, which may easily escape traditional experimental searches for magnetism.
Another prospective direction is to investigate the possibility of topological superconductivity, when both SOC and altermagnetism are present \cite{Li24,Ghorashi24,Zhu23},in particular in consideration of the staggered  DMI emerging in Sr$_2$RuO$_4$ when including SOC. Indeed, the additional spin component induced by the SOC may allow additional channels for the superconducting pairing.
Additionally, in terms of uncovering the origin of superconductivity in Sr$_2$RuO$_4$, it would be interesting to explore if superconductivity may emerge from altermagnetic fluctuations, and then what such fluctuations may give as the pairing symmetry \cite{mazin2022notes}. 
Our tight-binding model may also be used to extend the studies on superconductivity due to fluctuating loop currents in the case of distorted octahedra \cite{doi:10.1126/sciadv.adn3662,Mazzola2024}.

Finally, bulk Sr$_2$RuO$_4$ belongs to space group 64, which is the same as the parent cuprate compound La$_2$CuO$_4$. Therefore, the tight-binding model presented in this work for altermagnetism in Sr$_2$RuO$_4$ can also be used for hole-doped cuprates, simply by replacing the Ru $4d$ $xy$ orbitals with Cu $3d$ $x^2-y^2$ orbitals. These similarities, and the proposal that La$_2$CuO$_4$ also shows altermagnetism \cite{Smejkal22}, give shared properties to these two seemingly very different unconventional superconductors, and might provide answers to the longstanding puzzle of the origin of high-temperature superconductivity.

\section{Concluding remarks}
Lattice vibrations in Sr$_2$RuO$_4$ easily induce octahedral rotations. Given that Sr$_2$RuO$_4$ is also on the verge of a magnetic instability, such lattice distortions may induce interesting magnetic properties. In this work, we establish an orbital-selective altermagnetic state in Sr$_2$RuO$_4$ with finite octahedral rotation and establish its consequences both for the magnetic and superconducting properties. 

Using ab-initio calculations, we find for single layer Sr$_2$RuO$_4$ with octahedral rotation an orbital-selective $g$-wave altermagnetic phase. We further provide a low-energy $t_{2g}$ tight-binding model demonstrating that the $g$-wave altermagnetism predominantly arises as the product of the second and third nearest neighbor hybridizations between the $\gamma z$ ($\gamma=x,y$) orbitals, while the $xy$ orbitals only experience altermagnetism through a seventh nearest neighbor intraorbital process.
By replacing $xy$ with x$^2$-y$^2$ orbital, a similar tight-binding model could be used to investigate hole-doped cuprates within the same space group. 
In bulk Sr$_2$RuO$_4$ we establish the altermagnetic phase as the ground state for a reasonable range of octahedral rotation angles and Coulomb interactions. However, in the bulk, even the small interlayer hybridization breaks the symmetries of the $g$-wave altermagnet, driving a transition to a $d$-wave altermagnet. 
Finally, also including relativistic effects, we find that the spin-orbit coupling in Sr$_2$RuO$_4$ generates a staggered Dzyaloshinskii-Moriya interaction (DMI) responsible for weak ferromagnetism due to the N\'eel vector being in the $ab$ plane. However, if the N\'eel vector is rotated out-of-plane, ferromagnetism is absent.

The altermagnetic state of Sr$_2$RuO$_4$ is more complicated than many of the simplified altermagnet models proposed until now, which are often based on a single band picture. In particular, the $\gamma z$ orbitals in Sr$_2$RuO$_4$ experience altermagnetism through an interorbital third nearest neighbor hybridization generated by the octahedral rotation and with operator $\sigma_x^{orbital}$. Another interorbital second nearest neighbor hybridization (but inherent in unperturbed Sr$_2$RuO$_4$) is also needed to generate the higher $g$-wave altermagnetic symmetry. In contrast, the $xy$ orbitals only experience altermagnetism through much longer-range intraorbital hybridization. Overall, this leads to a highly orbital-selective altermagnetic state based on both inter- and intraorbital longer-range hybridization.
Interestingly, our results establish that the breaking of time-reversal symmetry in the altermagnetic state is notably different between the $\gamma z$ and $xy$ orbitals. This may agree with the experimental observation of time-reversal symmetry breaking in the $\alpha$ and $\beta$ Fermi surfaces\cite{PhysRevLett.83.3320}, which are composed of the $\gamma z$ orbitals. 

We further note that, while our results are obtained within the single layer and bulk Sr$_2$RuO$_4$, they can, with some modifications (such as the inclusion of interlayer DMI), be extended to the Sr$_2$RuO$_4$ surface. In fact, the surface of Sr$_2$RuO$_4$ is already known to spontaneously host large static octahedral distortions, $\phi>7^\circ$ \cite{Damascelli00,https://doi.org/10.1002/adma.202100593,Kreisel2021,Chandrasekaran2024}, likely favored by surface-enhanced electronic correlations and breaking of the crystal symmetry. Interestingly, a net magnetic moment along the $c$-axis, lower than 0.01~$\mu_B$ per Ru atom and with onset temperature greater than 50~K, has been experimentally reported on the surface \cite{Fittipaldi2021,Mazzola2024}. Theoretical explanations for this surface magnetism have so far been given in terms of orbital loop currents producing staggered magnetism as a consequence of the in-plane octahedral rotations \cite{Fittipaldi2021}. Our work instead points to the intriguing possibility that the octahedral rotations give rise to surface altermagnetism.

We also analyze some of the consequences of altermagnetism on the superconducting state in Sr$_2$RuO$_4$. Due to the strong spin-sublattice coupling in the altermagnetic phase and the resulting spin-splitting easily competing with the size of the superconducting gap, altermagnetism puts some restrictive constraints on the possible superconducting pairing symmetries. Notably, for spin-singlet intraorbital and in-plane pairing altermagnetism favors $d_{x^2-y^2}$- or $g$-wave symmetries, or time-reversal breaking or preserving combinations thereof, while $s$- or $d_{xy}$-wave symmetries even become forbidden for strong enough altermagnetism. Similarly, the simplest spin-triplet $p$-wave states are also not allowed. These results agree well with recent lists of pairing symmetry candidates based on decades of experimental measurements \cite{Maeno2024}. Our results are therefore compelling for narrowing down pairing symmetry possibilities in Sr$_2$RuO$_4$. We further note that our ab-initio results and the derived tight-binding model provide the necessary information to study the superconducting state and its interplay with the altermagnet phase within a realistic model that takes into account the multiorbital structure of Sr$_2$RuO$_4$, and thus additionally allows for the possibility of interorbital pairing. Within this framework, the effect of SOC on superconductivity can also be properly included for a more realistic description of the ruthenates.

\begin{acknowledgments}
We~acknowledge M.~Cuoco, A.~Damascelli, P.~Wahl, J.~B.~Profe and Y.~Liu for useful discussions.
This research was supported by the "MagTop" project (FENG.02.01-IP.05-0028/23) carried out within the "International Research
Agendas" programme of the Foundation for Polish Science, co-financed by the
European Union under the European Funds for Smart Economy 2021-2027 (FENG). ABS acknowledges support from the Knut and Alice Wallenberg Foundation KAW 2019.0309 through the Wallenberg Academy Fellows program and the Swedish Research Council (Vetenskapsr\aa det) grant agreement no.~2022-03963. PG acknowledges financial support from the Italian Ministry of University and Research (MUR) under the National Recovery and Resilience Plan (NRRP), Call PRIN 2022, funded by the European Union NextGenerationEU, Mission 4, Component 2, Grant No.2022HTPC2B
(TOTEM)- CUP B53D23004210006 and from NRRP MUR project PE0000023-NQSTI. GC acknowledges support from PNRR MUR project PE0000023-NQSTI.
We further acknowledge access to the computing facilities of the Interdisciplinary Center of Modeling at the University of Warsaw, Grant g96-1808 and g96-1809 for the availability of high-performance computing resources and support. We acknowledge the CINECA award under the ISCRA initiative IsC105 "SILENTSG" and IsB26 "SHINY" grants for the availability of high-performance computing resources and support. We acknowledge the access to the computing facilities of the Poznan Supercomputing and Networking Center Grant No. pl0267-01 and pl0365-01.
We acknowledge resources provided by the National Academic Infrastructure for Supercomputing in Sweden (NAISS), partially funded by the Swedish Research Council through grant agreement no.~2022-06725.
\end{acknowledgments}

\appendix

\begin{figure*}[htb]
\centering
\includegraphics[width=0.9\columnwidth,angle=0]{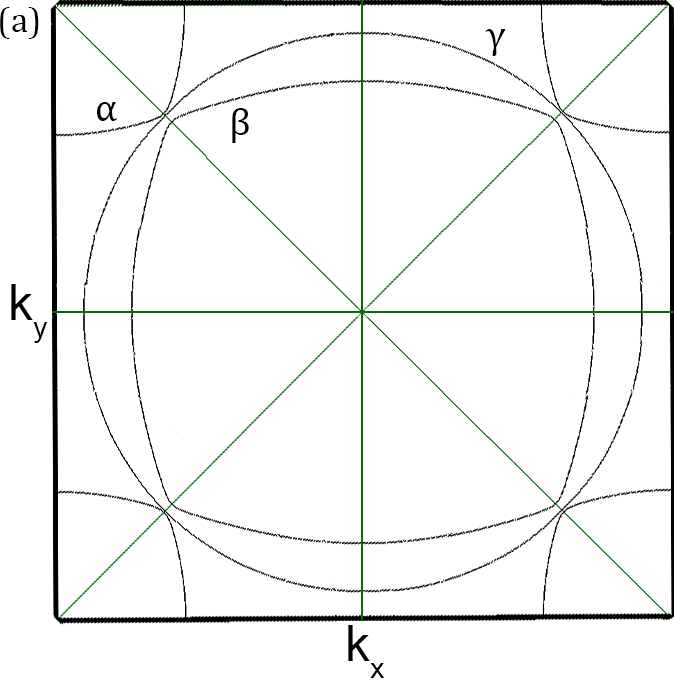}
\includegraphics[width=0.9\columnwidth,angle=0]{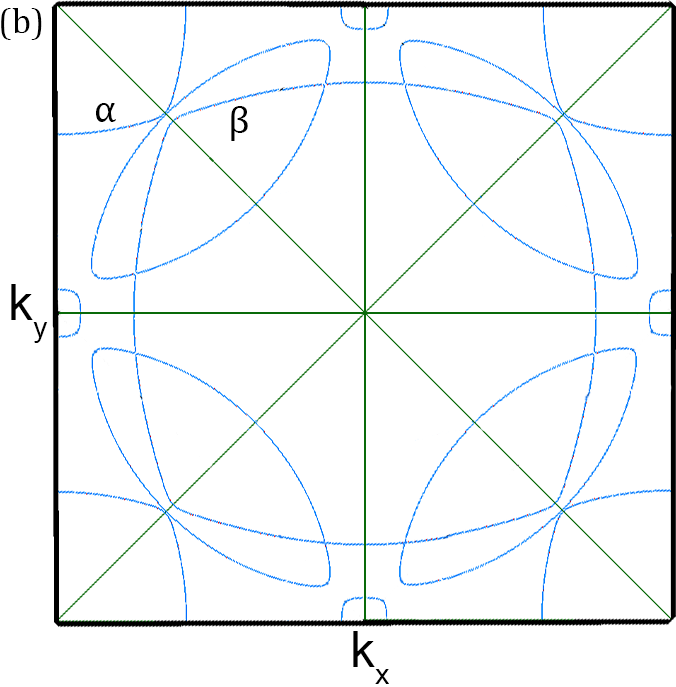}
\caption{Unfolded Fermi surfaces in extended Brilloiun zone for single layer Sr$_2$RuO$_4$ with $\phi=0.6^\circ$ and $U=0$ in the non-relativistic limit with no SOC. (a) Non-magnetic Fermi surface with $\alpha$, $\beta$ and $\gamma$ Fermi surfaces indicated. (b) Altermagnetic phase with a magnetic moment of 0.08 $\mu_B$ per atom, resulting in spin-up bands (blue) and spin-down bands (red) differing by less than 1 meV (not visible). The nodes of the Fermi surface along $k_x=0$, $k_y=0$, and $k_x=\pm k_y$ are indicated by dashed green lines.}\label{FS_unfolded}
\end{figure*}

\section{Unfolded and folded Fermi surfaces for single layer Sr$_2$RuO$_4$}
\label{app:FS}
In this Appendix, we provide additional information showing how the Fermi surfaces in Fig.~\ref{FS_singlelayer10} of the main text is related to the well-known Fermi surfaces of unperturbed bulk (and single layer) Sr$_2$RuO$_4$ \cite{Damascelli00,Bergemann03, PhysRevLett.116.106402,Tamai19, Maeno24a}.

We start by performing a calculation of the Fermi surface for single layer Sr$_2$RuO$_4$ with a small in-plane rotation angle of $\phi=0.6^\circ$ and at $U=0$. Due to the finite $\phi$ producing a doubling of the unit cell, hence halving of the Brillouin zone along with a $45^\circ$ rotation in the $k_x - k_y$ plane, we need to unfold and rotate the bands to be able to present the Fermi surfaces in the Brillouin zone of the conventional unit cell usually used in the literature \cite{Damascelli00,Bergemann03, PhysRevLett.116.106402,Tamai19, Maeno24a}.
In Fig.~\ref{FS_unfolded}(a), we present the Fermi surfaces in the unfolded, conventional Brillouin zone. 
As seen, the small in-plane octahedral rotations preserve the $\alpha$, $\beta$ and $\gamma$ Fermi surfaces. Here, the $\alpha$ band is centered around the zone corners, while the $\beta$ and $\gamma$ bands are centered around $\Gamma$, with the $\gamma$ band closest to the $\alpha$ band. Next, in Fig.~\ref{FS_unfolded}(b) we also allow for finite magnetic order, resulting in an altermagnetic phase. This affects the $\gamma$ band, which now forms Fermi pockets in the conventional Brillouin zone. 
A similar Fermi surface is also present in the antiferromagnetic phase. This Fermi surface effect is visible in the range $T_c<T<T_N$, where $T_c$ is the superconducting transition temperature and $T_N$ is the N\'eel temperature. However, no such temperature range has so far been found in Sr$_2$RuO$_4$ under stress, disorder, or pressure \cite{Grinenko2021,Grinenko2021phys}, but instead time-reversal symmetry breaking is found simultaneously with superconductivity. However, the $\gamma$ Fermi pockets, which may form arcs if not fully resolved, may be present in the cuprates \cite{PhysRevB.81.214524,Sachdev_2012}.

\begin{figure*}[htb]
\centering
\includegraphics[width=1.03\columnwidth,angle=0]{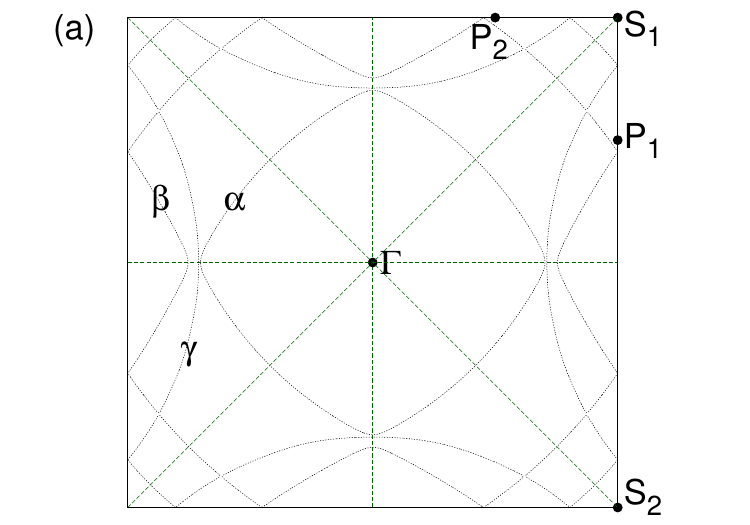}
\includegraphics[width=1.03\columnwidth,angle=0]{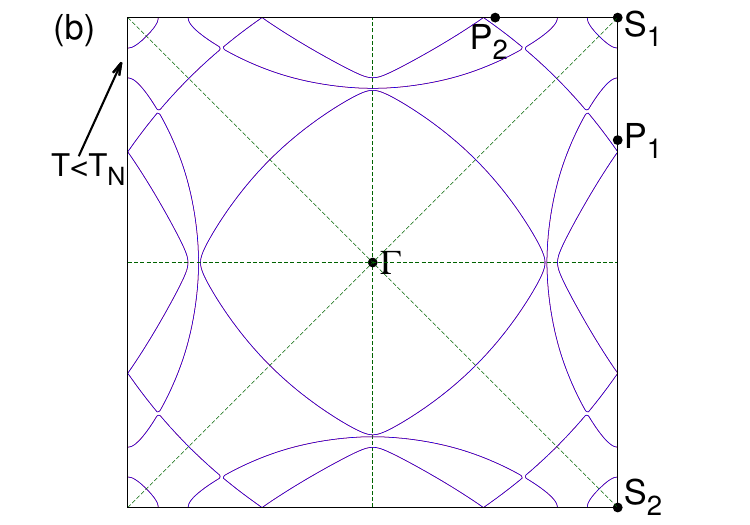}
\caption{Same as Fig.~\ref{FS_unfolded} but now in primitive Brillouin zone resulting in folded Fermi surfaces for $\phi=0.6^\circ$ and $U=0$. (a) Non-magnetic Fermi surfaces reported in black with $\alpha$, $\beta$ and $\gamma$ Fermi surfaces indicated. (b) Altermagnetic phase with a magnetic moment of 0.08 $\mu_B$ per atom, resulting in spin-up bands (blue) and spin-down bands (red) differing by less than 1 meV (not visible). The arrow indicates the only point of the Fermi surface affected by magnetism. The nodes of the Fermi surface along $k_x=0$, $k_y=0$, and $k_x=\pm k_y$ are indicated by dashed green lines.}\label{FS_folded}
\end{figure*}

Next, we plot the Fermi surfaces of Fig.~\ref{FS_unfolded} in the altermagnet unit cell in Fig.~\ref{FS_folded}, which then presents the folded and rotated bands of Fig.~\ref{FS_unfolded}. 
In the non-magnetic phase in Fig.~\ref{FS_folded}(a), we observe the folded bulk-like $\alpha$, $\beta$ and $\gamma$ Fermi surfaces, since the small rotation angle does not influence the topology of the Fermi surface.  In the magnetic phase in Fig.~\ref{FS_folded}(b), we see how the $\gamma$ band modification into Fermi pockets produces a change at the border of the Brillouin zone, indicated by an arrow, while the $\alpha$ and $\beta$ Fermi surfaces are unchanged. We further note that for $\phi=0.6^\circ$, there is also a Fermi pocket around the $S_1$ $k$-point, but this Fermi surface gaps out for larger values of $\phi$, otherwise, Fig.~\ref{FS_folded} closely reproduces the results in Fig.~\ref{FS_singlelayer10} of the main text.

For the small $\phi=0.6^\circ$ and $U$ used in Figs.~\ref{FS_unfolded}-\ref{FS_folded}, the non-relativistic spin-splitting is less than 1~meV. This is of the same order of magnitude as the superconducting gap in Sr$_2$RuO$_4$ and is therefore highly relevant when considering the low-temperature properties of Sr$_2$RuO$_4$. However, it is too small to appreciate as a spin-splitting on the Fermi surface visually. In this work, we therefore often study systems with the larger angle $\phi=8^\circ$ when plotting Fermi surfaces and band structures to be able to emphasize the altermagnetic symmetries even if it produces slightly modified Fermi surfaces.

\section{Additional band structure plots for single layer Sr$_2$RuO$_4$}
\label{app:band}
\begin{figure}[htb]
\centering
\includegraphics[width=5.9cm,height=\columnwidth,angle=270]{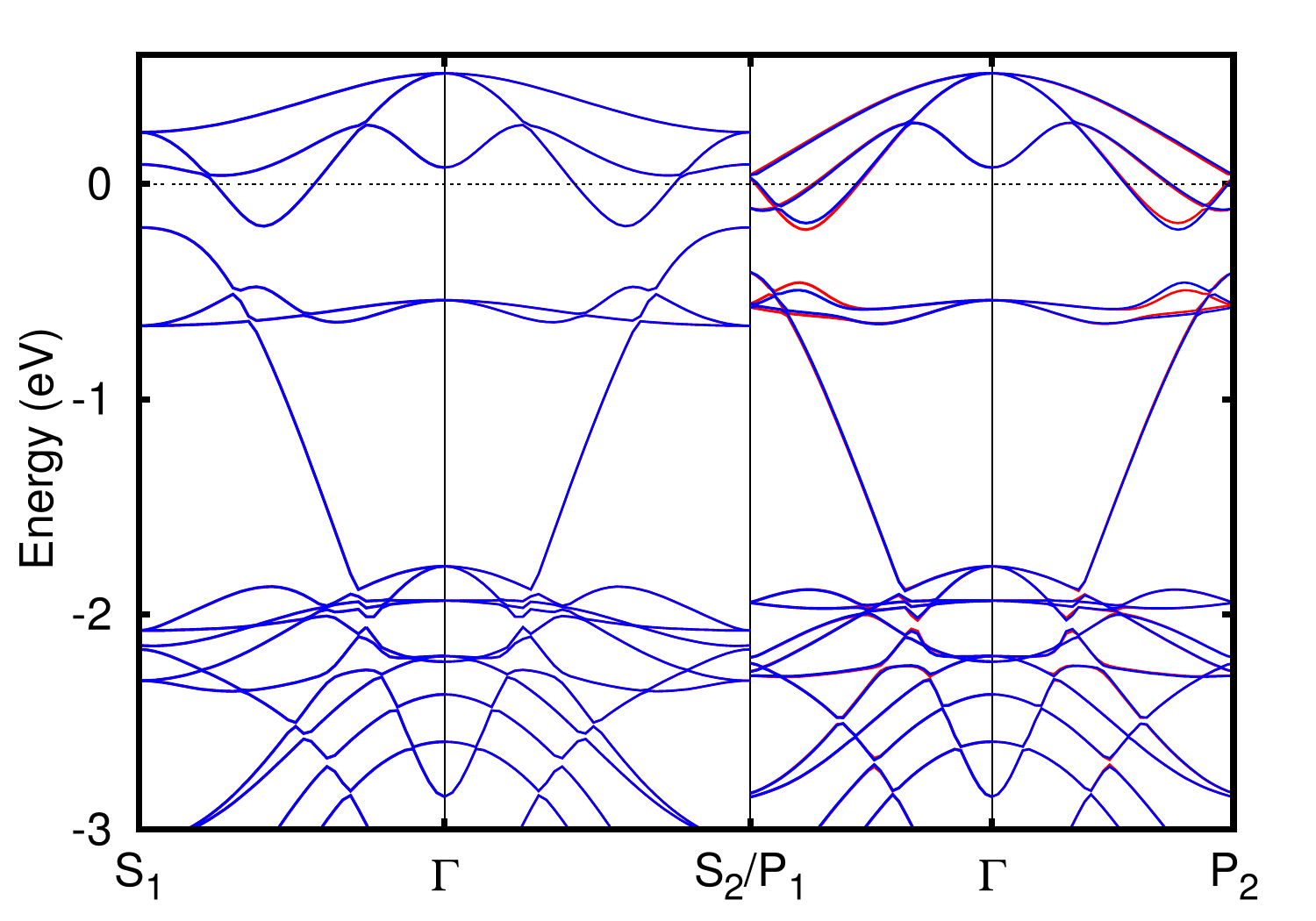}
\caption{Band structure of the single layer of Sr$_2$RuO$_4$ for $U=1.0$~eV and $\phi$ =8$^{\circ}$, with spin-up bands (blue) and spin-down bands (red) in the non-relativistic limit with no SOC. Similar to Fig.~\ref{BS_singlelayer} in the main text, but here over an extended energy range.
}\label{BS_singlelayer_narrow}
\end{figure}

In this Appendix, we report two additional band structure plots for the single layer Sr$_2$RuO$_4$ to supplement the discussion in Section \ref{sec:3}.
First, in Fig.~\ref{BS_singlelayer_narrow} we complement Fig.~\ref{BS_singlelayer} in the main text by plotting the band structure over an extended energy range, which then also includes the full $xy$-derived band, as well as oxygen-derived bands below $-2$~eV, as also characteristic of the non-magnetic bulk Sr$_2$RuO$_4$\cite{PhysRevB.89.075102}. As seen, no spin-splitting is present on the $xy$-derived band at negative energies and it is also negligible on the oxygen-derived bands. This verifies that the associated tight-binding model primarily needs to focus on the $\gamma z$ orbitals. 

\begin{figure}[htb]
\centering
\includegraphics[width=5.9cm,height=\columnwidth,angle=270]{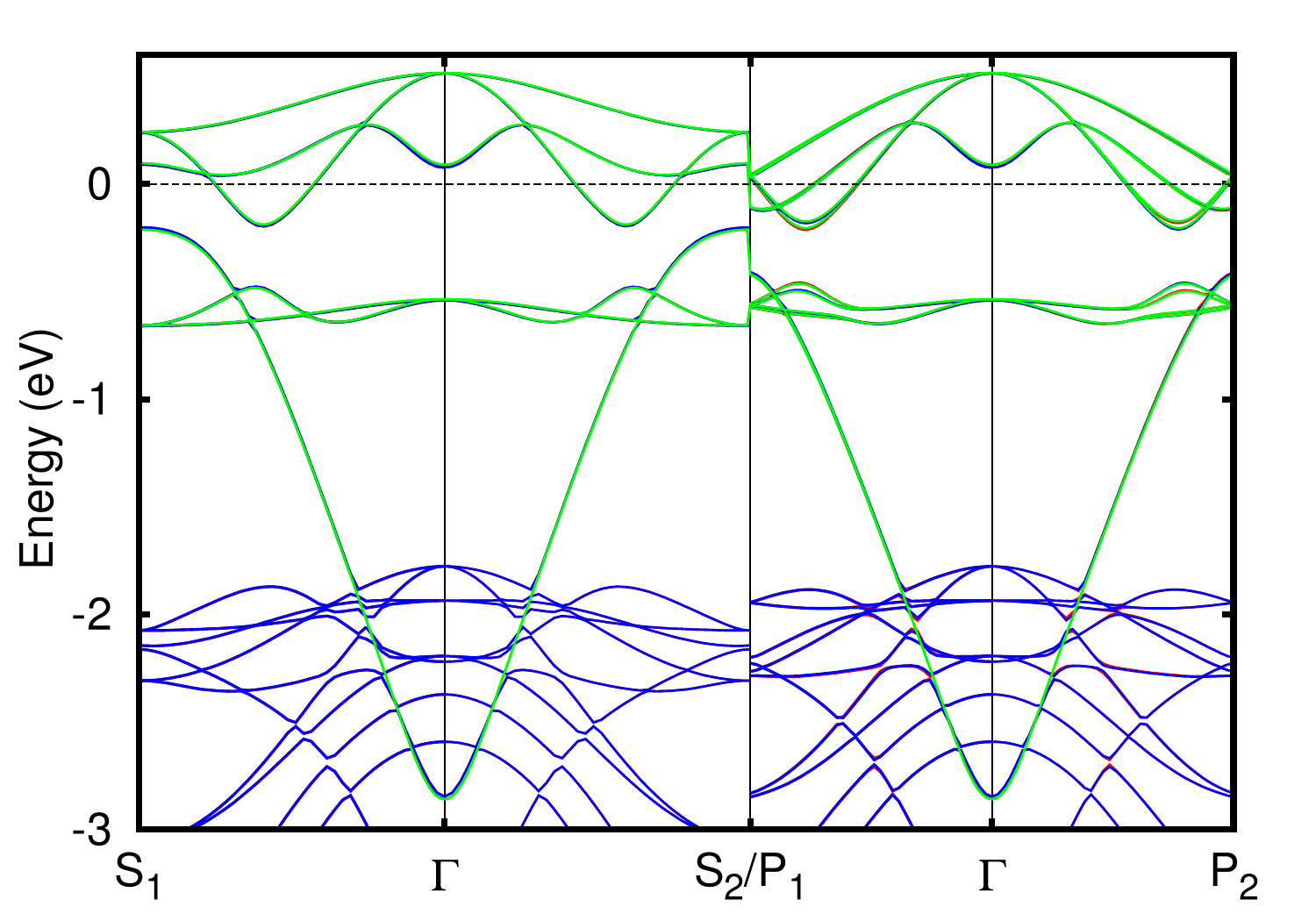}
\caption{Band structure of the single layer Sr$_2$RuO$_4$ for $U=1.0$~eV and $\phi=8^{\circ}$, with spin-up bands (blue), spin-down bands (red) in the non-relativistic limit with no SOC, and the band structure obtained by wannierization of the $t_{2g}$ electrons (green).}
\label{BS_singlelayer_wannier}
\end{figure}

To extract an accurate Fermi surface, we perform a wannierization. We do so for the subsectors with spin-up and spin-down spins from the same self-consistent calculation, after imposing zero net magnetization and without spin-orbit coupling. The band structure obtained from the wannierization is reported in Fig.~\ref{BS_singlelayer_wannier}, overlayed over the ab-initio band structure, showing an overall excellent fit, which verifies that the extracted Fermi surfaces will be of high quality.
In general, the wannierization is very good, but it becomes somewhat worse in the regions of $k$-space where the non-relativistic spin-splitting is maximum. Using the maximally localization procedure helps in the fitting of the band structure, also in these $k$-space regions with large non-relativistic spin-splittings.

\section{Tight-binding model for single layer Sr$_2$RuO$_4$ $g$-wave altermagnetic phase}
\label{app:TB}

In this Appendix, we derive the full low-energy Hamiltonian in the altermagnet phase for single layer Sr$_2$RuO$_4$, including all relevant low-energy orbitals, which are the Ru $t_{2g}$ orbitals, and including up to third NN processes. Apart from the octahedral distortion and the doubling of the expanded unit cell, the model contains the main hopping already reported in previous works \cite{PhysRevB.74.035115,PhysRevB.89.075102}.
We already at the outset note that, while previous Hamiltonian models for altermagnetic systems have usually been obtained by adding to the non-magnetic band structure an intraband hopping with a Pauli matrix $\sigma_z$ operator with respect to the site \cite{Smejkal22beyond}, instead, for Sr$_2$RuO$_4$ and by extension also for related materials, we establish how an interorbital hopping parameters between $\gamma z$ bands is the key enabler for altermagnetism.
As a consequence, we need to explicitly keep this orbital band, or equivalently degree of freedom.

\begin{figure*}
\centering
\includegraphics[width=17cm,angle=0]{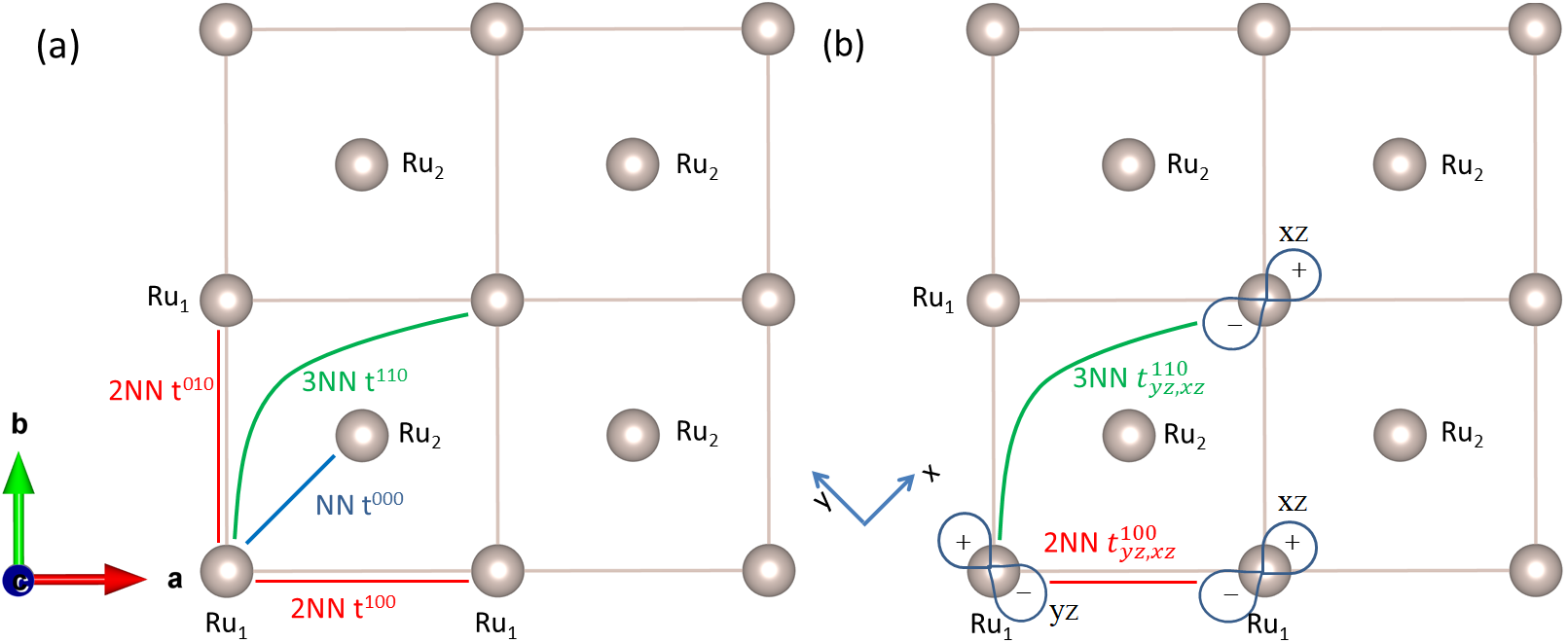}
\caption{Top view of the schematic representation of single-layer Sr$_2$RuO$_4$ lattice with the Ru$_1$ and Ru$_2$ atoms (grey) within a 2$\times$2 repetition of the primitive unit cell in the ab plane, while the oxygen and strontium atoms are not shown. In addition, (a) shows first, second, and third NNs hoppings represented by blue, red and green lines, respectively, corresponding to the terms  $t^{000}$, $t^{100}$ and $t^{110}$. (b) Reports the original (primitive unit cell) reference system for the $t_{2g}$ orbitals with the $x$- and $y$-directions along the NN Ru-Ru directions. Bottom left corner atom shows a top view of the $yz$ orbital, while the other two atoms display two $xz$ orbitals. In the absence of in-plane octahedral rotations, the third NN hopping (green) t$^{110}_{yz,xz}=0$ by symmetry, while the second NN (red) t$^{100}_{yz,xz}$ is always finite.}
\label{figure9}
\end{figure*}

To capture any putative altermagnet phase we have an extended unit cell with two Ru sites: Ru$_1$ at coordinates $(0,0,0)$ and Ru$_2$ at coordinates $(0.5,0.5,0)$, where we measure in units of the extended unit cell with size $a = \sqrt{2}a_{uc}$, see Fig.~\ref{figure9}, with $a_{uc}=3.86$~\AA \cite{PhysRevB.57.5067}. For simplicity, we set $a=1$ in all trigonometric functions. 
We use the orbital basis composed of (Ru$_1$ $d_{xy\uparrow}$, Ru$_1$ $d_{xz\uparrow}$, Ru$_1$ $d_{yz\uparrow}$, Ru$_2$ $d_{xy\uparrow}$, Ru$_2$ $d_{xz\uparrow}$, Ru$_2$ $d_{yz\uparrow}$) and equivalently for the spin-down sector. In total, this results in a  12$\times$12 Hamiltonian: 
\begin{align}
\label{eq:HTB}
 \hat{H}(k_x,k_y) = \left( \begin{array}{cccc}
H^{\uparrow\uparrow} & H^{\uparrow\downarrow} \\
H^{\downarrow\uparrow} & H^{\downarrow\downarrow} \\
 \end{array}\right), 
 \end{align}
where the 6$\times$6 Hamiltonian $H^{\uparrow\uparrow}$ in terms of the Ru atom 1 and atom 2 reads:
\begin{align}
\label{eq:HTB2}
H^{\uparrow\uparrow} = \left( \begin{array}{cccc}
H_{11}^{\uparrow\uparrow} & H_{12}^{\uparrow\uparrow} \\
H_{21}^{\uparrow\uparrow} & H_{22}^{\uparrow\uparrow} \\
\end{array}\right), 
\end{align}
 and similarly in the other spin sectors.
The terms $H_{11}^{\uparrow\uparrow}$ and $H_{22}^{\uparrow\uparrow}$ contain on-site energies as well as second and third NN hopping terms, which we present in detail below. 
The $H_{12}^{\uparrow\uparrow}$ term contains the NN hopping necessary for a realistic description of the band structure; but is not relevant for altermagnetism within our approximations as it is not affected to first order by the octahedral distortion. 

All the Ru atoms are equivalent; therefore, they have the same energies on site.
To start, we write down the matrix elements for the spin-up Hamiltonian subsector. Including onsite energies $\varepsilon^0$ as well as first, second, and third NN hopping parameters $t$, also illustrated in Fig.~\ref{figure9}(a), we arrive at:
\begin{widetext}
\[
\scalebox{0.92}{$
H_{11}^{\uparrow\uparrow} = \left( \begin{smallmatrix}
\varepsilon^0_{xy,xy} + 2t^{100}_{1xy,1xy}(\cos{k_x} + \cos{k_y})+\textcolor{purple}{r(k_x,k_y)} & 0 & 0 \\
0 & \varepsilon^0_{xz,xz} + 2t^{100}_{1xz,1xz}\cos{k_x} + 2t^{010}_{1xz,1xz}\cos{k_y} & 2t^{100}_{1xz,1yz}\cos{k_x} - 2t^{010}_{1xz,1yz}\cos{k_y} + \textcolor{green}{4t^{110}_{xz,yz}\sin{k_x}\sin{k_y}} \\
0 & 2t^{100}_{1xz,1yz}\cos{k_x} - 2t^{010}_{1xz,1yz}\cos{k_y} + \textcolor{green}{4t^{110}_{xz,yz}\sin{k_x}\sin{k_y}} & \varepsilon^0_{yz,yz} + 2t^{100}_{1yz,1yz}\cos{k_x} + 2t^{010}_{1yz,1yz}\cos{k_y}
\end{smallmatrix} \right)
$}
\]

\[
\scalebox{0.92}{$
H_{22}^{\uparrow\uparrow} = \left( \begin{smallmatrix}
\varepsilon^0_{xy,xy} + 2t^{100}_{2xy,2xy}(\cos{k_x} + \cos{k_y}) -\textcolor{purple}{r(k_x,k_y)} & 0 & 0 \\
0 & \varepsilon^0_{xz,xz}  + 2t^{100}_{2xz,2xz}\cos{k_x} + 2t^{010}_{2xz,2xz}\cos{k_y} & 2t^{100}_{2xz,2yz}\cos{k_x} - 2t^{010}_{2xz,2yz}\cos{k_y} - \textcolor{green}{4t^{110}_{xz,yz}\sin{k_x}\sin{k_y}} \\
0 & 2t^{100}_{2xz,2yz}\cos{k_x} - 2t^{010}_{2xz,2yz}\cos{k_y} - \textcolor{green}{4t^{110}_{xz,yz}\sin{k_x}\sin{k_y}} & \varepsilon^0_{yz,yz} + 2t^{100}_{2yz,2yz}\cos{k_x} + 2t^{010}_{2yz,2yz}\cos{k_y}
\end{smallmatrix} \right)
$}
\]
\[ H_{12}^{\uparrow\uparrow} = \left( \begin{smallmatrix}
4t^{000}_{1xy,2xy}\cos{k_x}\cos{k_y} & 0 & 0 \\
0 & 2t^{\frac{1}{2}\frac{1}{2}0}_{1xz,2xz}\cos{(k_x+k_y)/2} +2t^{-\frac{1}{2}\frac{1}{2}0}_{1xz,2xz}\cos{(k_x-k_y)/2} & 0 \\
0 & 0 & 2t^{\frac{1}{2}\frac{1}{2}0}_{1yz,2yz}\cos{(k_x+k_y)/2} +2t^{-\frac{1}{2}\frac{1}{2}0}_{1yz,2yz}\cos{(k_x-k_y)/2}
\end{smallmatrix} \right). \]
\end{widetext}
Here we have included the standard terms up to third NN interactions in black and also added the interorbital contributions arising from a finite octahedral distortion as terms in green, matching the color in Fig.~\ref{figure9}. These interorbital contributions are from a third NN hopping process, which is only non-zero when the lattice is distorted, as seen directly in Fig.~\ref{figure9}. 
Moreover, for completeness, we have also included an intraorbital term $r(k_x,k_y)=8t^{210}_{xy,xy}\sin{k_y}\sin{k_y}(\cos{k_x}-\cos{k_y})$, indicated in purple. This is a seventh NN hopping contribution and the largest contribution (i.e.~shortest hopping distance) allowed by symmetry that may generate $g$-wave altermagnetism in the ${xy}$ orbitals. Due to its long-range nature, it is generally smaller than the altermagnetic contributions we derive for the $\gamma z$ orbitals. However, at the Fermi surface this non-relativistic spin-splitting can also become reasonably large.

The Hamiltonians above can be substantially simplified by identifying symmetry relations between the different parameters.
The on-site terms $\varepsilon^0$ can, due to symmetry, further be simplified as 
\begin{align}
\varepsilon_{xz} & \coloneq \varepsilon^0_{xz,xz}= \varepsilon^0_{yz,yz} \, \,
{\rm and, likewise,} \nonumber \\ 
\varepsilon_{xy} & \coloneq \varepsilon^0_{xy,xy} \nonumber
\end{align}
For all the hopping parameters, we can also simplify the notation using the following symmetry-enforced relations for the $xy$ hoppings: 
\begin{align}
t^{100}_{xy,xy} & \coloneq t^{100}_{1xy,1xy}=t^{010}_{1xy,1xy}; \nonumber \\ \nonumber
t^{100}_{xy,xy} & \coloneq t^{100}_{2xy,2xy}=t^{010}_{2xy,2xy}; \\ \nonumber
\end{align}
and for the  ${\gamma}z$ hoppings:
\begin{align}
t^{100}_{xz,xz} & \coloneq t^{100}_{1xz,1xz}=t^{010}_{1yz,1yz}; \\ \nonumber
t^{010}_{xz,xz} & \coloneq t^{010}_{1xz,1xz}=t^{100}_{1yz,1yz}; \\ \nonumber
t^{100}_{xz,xz} & \coloneq t^{100}_{2xz,2xz}=t^{010}_{2yz,2yz};\\ \nonumber
t^{010}_{xz,xz} & \coloneq t^{010}_{2xz,2xz}=t^{100}_{2yz,2yz}.\\ \nonumber
\end{align}
In preparation for a putative finite spin-splitting, we further introduce on-site spin-splittings $\Delta_{xy}$ and $\Delta_{{\gamma}z}$ for the $xy$ and ${\gamma}z$ orbitals, respectively, colorcoded blue in the following. Here we encode the Ru$_1$  majority channel as the spin $\uparrow$ channel, while for Ru$_2$ the majority channel is the spin $\downarrow$ channel.
To summarize, we arrive at:
\begin{widetext}
\begin{align}
\label{eq:H11}
H_{11}^{\uparrow\uparrow} = \left( \begin{smallmatrix}
\varepsilon_{xy} -\textcolor{blue}{\frac{\Delta_{xy}}{2}} + 2t^{100}_{xy,xy}(\cos{k_x} +\cos{k_y})+\textcolor{purple}{r(k_x,k_y)} & 0 & 0 \\
0 & \varepsilon_{xz} -\textcolor{blue}{\frac{\Delta_{{\gamma}z}}{2}}+2t^{100}_{xz,xz}\cos{k_x} +2t^{010}_{xz,xz}\cos{k_y} &  2t^{100}_{xz,yz}(\cos{k_x}-\cos{k_y})+
\textcolor{green}{4t^{110}_{xz,yz}\sin{k_x}\sin{k_y}} \\
0 &  2t^{100}_{xz,yz}(\cos{k_x}-\cos{k_y})+
\textcolor{green}{4t^{110}_{xz,yz}\sin{k_x}\sin{k_y}} & \varepsilon_{xz} -\textcolor{blue}{\frac{\Delta_{{\gamma}z}}{2}}+2t^{010}_{xz,xz}\cos{k_x} +2t^{100}_{xz,xz}\cos{k_y}
\end{smallmatrix} \right)
\end{align}
\begin{align}
\label{eq:H22}
H_{22}^{\uparrow\uparrow} = \left( \begin{smallmatrix}
\varepsilon_{xy} +\textcolor{blue}{\frac{\Delta_{xy}}{2}}+2t^{100}_{xy,xy}(\cos{k_x} +\cos{k_y})  -\textcolor{purple}{r(k_x,k_y)}& 0 & 0 \\
0 & \varepsilon_{xz} +\textcolor{blue}{\frac{\Delta_{{\gamma}z}}{2}}+2t^{100}_{xz,xz}\cos{k_x} +2t^{010}_{xz,xz}\cos{k_y} & 2t^{100}_{xz,yz}(\cos{k_x}-\cos{k_y})-\textcolor{green}{4t^{110}_{xz,yz}\sin{k_x}\sin{k_y}} \\
0 &  2t^{100}_{xz,yz}(\cos{k_x}-\cos{k_y})-\textcolor{green}{4t^{110}_{xz,yz}\sin{k_x}\sin{k_y}} & \varepsilon_{xz} +\textcolor{blue}{\frac{\Delta_{{\gamma}z}}{2}}+2t^{010}_{xz,xz}\cos{k_x} +2t^{100}_{xz,xz}\cos{k_y}
\end{smallmatrix} \right),
\end{align}
\end{widetext}
Here terms in green mark the largest contributions from the octahedral distortions, which are induced by third NN interorbital hopping terms. No other terms are induced in the $t_{2g}$ subspace up to third NNs from octahedral distortion.
Terms in purple are also induced by the octahedral distortion, but only take place between seventh NNs. Despite our model only containing terms generically up to third NNs, we need to include this term as it is the closest-range hopping term we find that may generate altermagnetism in the $xy$ orbitals. Finally, blue terms are onsite energies that become finite if a spin-splitting were to be present in either the $xy$ or $\gamma z$ orbitals.  We further note that the Hamiltonian $H_{12}^{\uparrow \uparrow}$ contains no contribution from octahedral distortion or altermagnetism and thus remains unchanged from the earlier definition (up to the new definitions of $\varepsilon^0$ and $t$).  
The lower $2 \times 2$ diagonal blocks of Eqs.~\eqref{eq:H11} and \eqref{eq:H22} are reproduced as $H_{\gamma z} = H^0_{\gamma z} +H^{\rm AM}_{\gamma z}$ through Eqs.~\eqref{eq:H0} and \eqref{eq:HAM} in the main text and hence gives all components up to third NNs present in the altermagnetic state. Focusing on only the $\gamma z$ orbitals in the main text is allowed due to the diagonal block nature of Eqs.~\eqref{eq:H11}-\eqref{eq:H22}, which at this level do not include any coupling at the ${xy}$ to the $\gamma z$ orbitals. Although we find numerically in Fig.~\ref{FS_singlelayer10} that altermagnetism is also present in the $xy$-orbital, i.e.~on the $\gamma$ surface, its contribution is due to a longer-range intraorbital process, which therefore is generally smaller and also does not involve the $\gamma z$ orbitals.

Considering the Hamiltonian in equation (\ref{eq:HTB2}), the spin-down subsector can be obtained using the property:
\begin{equation} \label{A3}
H^{\downarrow\downarrow}(k_x,k_y,\Delta_{xy},\Delta_{{\gamma}z}) = H^{\uparrow\uparrow}(k_x,k_y,-\Delta_{xy},-\Delta_{{\gamma}z})
\end{equation}
and due to the symmetries of the crystal structure, we have also:
\begin{equation*} 
H^{\downarrow\downarrow}(k_y,k_x) = H^{\uparrow\uparrow}(k_x,k_y).
\end{equation*}
Furthermore, $H^{\uparrow\downarrow}$ and
$H^{\downarrow\uparrow}$ contain only spin-orbit terms, which we ignore in the non-relativistic limit and hence we here set $H^{\uparrow\downarrow} = H^{\downarrow\uparrow} = 0$. 

To be able to highlight the altermagnetic effect analytically, we simplify the Hamiltonian $\hat{H}$ by assuming the first NN hopping to be zero in $H_{12}^{\uparrow\uparrow}$, and thus likewise in $H_{12}^{\downarrow\downarrow}$. While this is unrealistic in order to reproduce the full band structure, it still helps to illustrate the altermagnetic effect. With these simplifications, we arrive at the following eigenvalues for the spin-up sector of the tight-binding Hamiltonian Eq.~\eqref{eq:HTB}:
\begin{equation*} \varepsilon(k_x,k_y)^{\uparrow\uparrow}_{11}=\varepsilon_{xz}-\frac{\Delta_{{\gamma}z}}{2} \pm \sqrt{g(k_x,k_y)_{+}}
\end{equation*}
\begin{equation*}
\varepsilon(k_x,k_y)^{\uparrow\uparrow}_{22}=\varepsilon_{xz}+\frac{\Delta_{{\gamma}z}}{2} \pm \sqrt{g(k_x,k_y)_{-}},
\end{equation*}
where
$g(k_x,k_y)_{\pm}=f(k_x,k_y)\pm A(\cos{k_x}-\cos{k_y})\sin{k_x}\sin{k_y}$, with
$f(k_x,k_y)$ being a function and $A$ a finite coefficient related to the hopping parameters. 
Furthermore, from Eq.~(\ref{A3}), we obtain $\varepsilon(k_x,k_y,\Delta_{xy},\Delta_{{\gamma}z})^{\downarrow\downarrow}_{11}=\varepsilon(k_x,k_y,-\Delta_{xy},-\Delta_{{\gamma}z})^{\uparrow\uparrow}_{22}$. We thus find that $\varepsilon(k_x,k_y)^{\downarrow\downarrow}_{11}$ and
$\varepsilon(k_x,k_y)^{\uparrow\uparrow}_{11}$ only differ because of the term $(\cos{k_x}-\cos{k_y})\sin{k_x}\sin{k_y}$. This term has the structure of a $g$-wave orbital: $(x^2-y^2)xy$, which is exactly the altermagnetic order obtained for the Fermi surface in Fig.~\ref{FS_singlelayer10} in the main text. This shows how the derived tight-binding model is able to describe an altermagnetic $g$-wave order of the kind $(x^2-y^2)xy$. 

Having derived the low-energy tight-binding model for the $g$-wave altermagnet structure, we next discuss the values of the ingoing parameters. 
In previous literature \cite{PhysRevB.74.035115,PhysRevB.89.075102}, the hopping parameters have usually been provided for a unit cell with 1 Ru atom. In order to map our notation with this previously used notation, we define $\tilde{t}$ to be the hopping parameters for the unit cell with 1 Ru atom. 
We also need to consider that the primitive unit cell's $x$-axis is along the direction of the vector of the supercell \textbf{a}+\textbf{b} direction, along the bond between the Ru$_1$ and Ru$_2$ atoms. Taking this into account, we have the following equalities with earlier derived parameters \cite{PhysRevB.89.075102} and for the on-site energy and the first-neighbor hopping we have:
\begin{equation}
\varepsilon_{xy}-\varepsilon_{xz}=-153 \,\, \textnormal{meV}, \nonumber
\end{equation}
\begin{equation}
t^{\frac{1}{2}\frac{1}{2}0}_{1xy,2xy}=\tilde{t}^{100}_{xy,xy}=-387 \, \,\textnormal{meV}, \nonumber
\end{equation}
\begin{equation}
t^{\frac{1}{2}\frac{1}{2}0}_{1xz,2xz}=t^{-\frac{1}{2}\frac{1}{2}0}_{1yz,2yz}=\tilde{t}^{100}_{xz,xz}=-291 \, \, \textnormal{meV}, \nonumber
\end{equation}
\begin{equation}
t^{-\frac{1}{2}\frac{1}{2}0}_{1xz,2xz}=t^{\frac{1}{2}\frac{1}{2}0}_{1yz,2yz}=\tilde{t}^{010}_{xz,xz}=-39 \,\, \textnormal{meV}, \nonumber
\end{equation}
The last two hopping parameters are different since they are associated with the $\pi$ and $\delta$ bonds, respectively \cite{PhysRevB.74.035115,PhysRevB.89.075102}.
For the second-neighbor hoppings, we have:
\begin{equation}
t^{100}_{xy,xy}=\tilde{t}^{110}_{xy,xy}=-138 \,\, \textnormal{meV}, \nonumber
\end{equation}
\begin{equation}
t^{100}_{xz,xz}=\tilde{t}^{110}_{xz,xz}=+17 \,\, \textnormal{meV}. \nonumber
\end{equation}
\begin{equation}
t^{010}_{xz,yz}=-t^{100}_{xz,yz}=\tilde{t}^{110}_{xz,yz}=-12 \,\, \textnormal{meV}, \nonumber
\end{equation}
where the last terms are the only non-zero interorbital hoppings, as seen from Fig.~\ref{figure9}(b). The effect of this hopping on the magnetic properties of ruthenates has already been investigated \cite{Vanthiel21}. 

Finally, we include one of the third NN hoppings, the shortest NN hopping that generates altermagnetism. The in-plane octahedral distortion generates this term:
\begin{equation}
t^{110}_{xz,yz}=\tilde{t}^{200}_{xz,yz}. \nonumber
\end{equation}
This hopping is zero in the undistorted unit cell ($\phi$=0), but with its absolute value increasing proportionally to sin($\phi$) with the rotation angle. 
For $\phi$=1$^\circ$, which is in the range of expected rotations, we have $t^{110}_{xz,yz}\approx 1$~meV, which is small but notably, even an infinitesimal value breaks time-reversal symmetry. 
Furthermore, for a rotation angle $\phi = 8^{\circ}$ and for $U=0.2$~eV, we extract the non-relativistic spin-splitting terms $\Delta_{xy} = -84$~meV and $\Delta_{{\gamma}z} = 68$~meV, while for $U=1.0$~eV, these even increase to $\Delta_{xy} = -320$~meV and $\Delta_{{\gamma}z} = 288$~meV, which are all very large. Finally, in Appendix \ref{app:magphase} we also show how an in-plane octahedral angle of  $\phi = 8^\circ$ produces an altermagnetic spin-splitting between 3-7~meV at the Fermi surface. For a more realistic and smaller $\phi \approx 1^\circ$, the spin-splittig would be smaller, but likely still comparable to the superconducting gap value of 350~${\mu}$eV \cite{doi:10.1073/pnas.1916463117}.

In this setup, we neglect the long-range terms $\tilde{t}^{200}_{xy,xy}$, $\tilde{t}^{200}_{xz,xz}$ etc., present in the undistorted case as they are not essential to capture the altermagnetic effect. We also neglect the term $\tilde{t}^{100}_{xz,yz}$ arising from an octahedral rotation, but not necessary in order to describe the $g$-wave altermagnetism. With the above choices of parameters, the tight-binding structure reported in this Appendix relatively well reproduces the band structure in Fig.~\ref{FS_singlelayer10}.

Finally, a few remarks are in order. First of all, moving beyond the single layer limit and considering also interlayer hopping parameters in the tight-binding model, we find that the interlayer hopping terms, which destroy the $g$-wave altermagnet order and produce the $d_{xy}$-wave altermagnet are the following \cite{PhysRevB.74.035115,PhysRevB.89.075102}:
\begin{equation}
t^{\frac{1}{2}0\frac{1}{2}}_{1xz,3xz}=\tilde{t}^{\frac{1}{2}\frac{1}{2}\frac{1}{2}}_{xz,xz}=
t^{\frac{1}{2}0\frac{1}{2}}_{1yz,3yz}=\tilde{t}^{\frac{1}{2}\frac{1}{2}\frac{1}{2}}_{yz,yz}=-19 \,\, \textnormal{meV}, \nonumber
\end{equation}
\begin{equation}
t^{\frac{1}{2}0\frac{1}{2}}_{1xz,3yz}=\tilde{t}^{\frac{1}{2}\frac{1}{2}\frac{1}{2}}_{xz,yz}=-14 \,\, \textnormal{meV},  \nonumber
\end{equation}
\begin{equation}
t^{\frac{1}{2}0\frac{1}{2}}_{1xz,3xy}=\tilde{t}^{\frac{1}{2}\frac{1}{2}\frac{1}{2}}_{xz,xy}= +6 \,\, \textnormal{meV}, \nonumber
\end{equation}
where the index 3 indicates the Ru$_3$ atom in the RuO$_2$ layer, different from the layer containing the Ru$_1$ and Ru$_2$ atoms.
We note that the interlayer hopping parameters are one order of magnitude smaller than intralayer hopping parameters in Sr$_2$RuO$_4$ \cite{PhysRevB.74.035115,PhysRevB.89.075102}. Still, we find that even an infinitesimal interlayer hopping breaks the $g$-wave order down to the $d_{xy}$-wave symmetry. Thus, even if most of the bulk Fermi surface seems to host a $g$-wave altermagntic order, from the mere mathematical point of view the small value of the interlayer hoppings produces a $d_{xy}$-wave solution. 

Second, regarding other possible structural distortions, we calculate that the Jahn-Teller (JT) effect without octahedral rotations also produces the $g$-wave altermagnet in single layer A$_2$BO$_4$, and having both JT and octahedral rotation still produces a $g$-wave altermagnetism in the single layer.
Finally, if we add also an octahedral tilting we move to space group 61 and the hybridization induces altermagnetism in the orbitals even with respect to the basal plane. However, the altermagnetism is still strongly suppressed in the orbitals that are not the $\gamma z$-orbitals, so we still can define this magnetic phase as an orbital-selective altermagnetism\cite{Cuono23orbital}.

\section{Magnetic phases as a function of octahedral rotation angle}
\label{app:magphase}
\begin{figure}[htb]
\centering
\includegraphics[width=5.9cm,height=\columnwidth,angle=270]{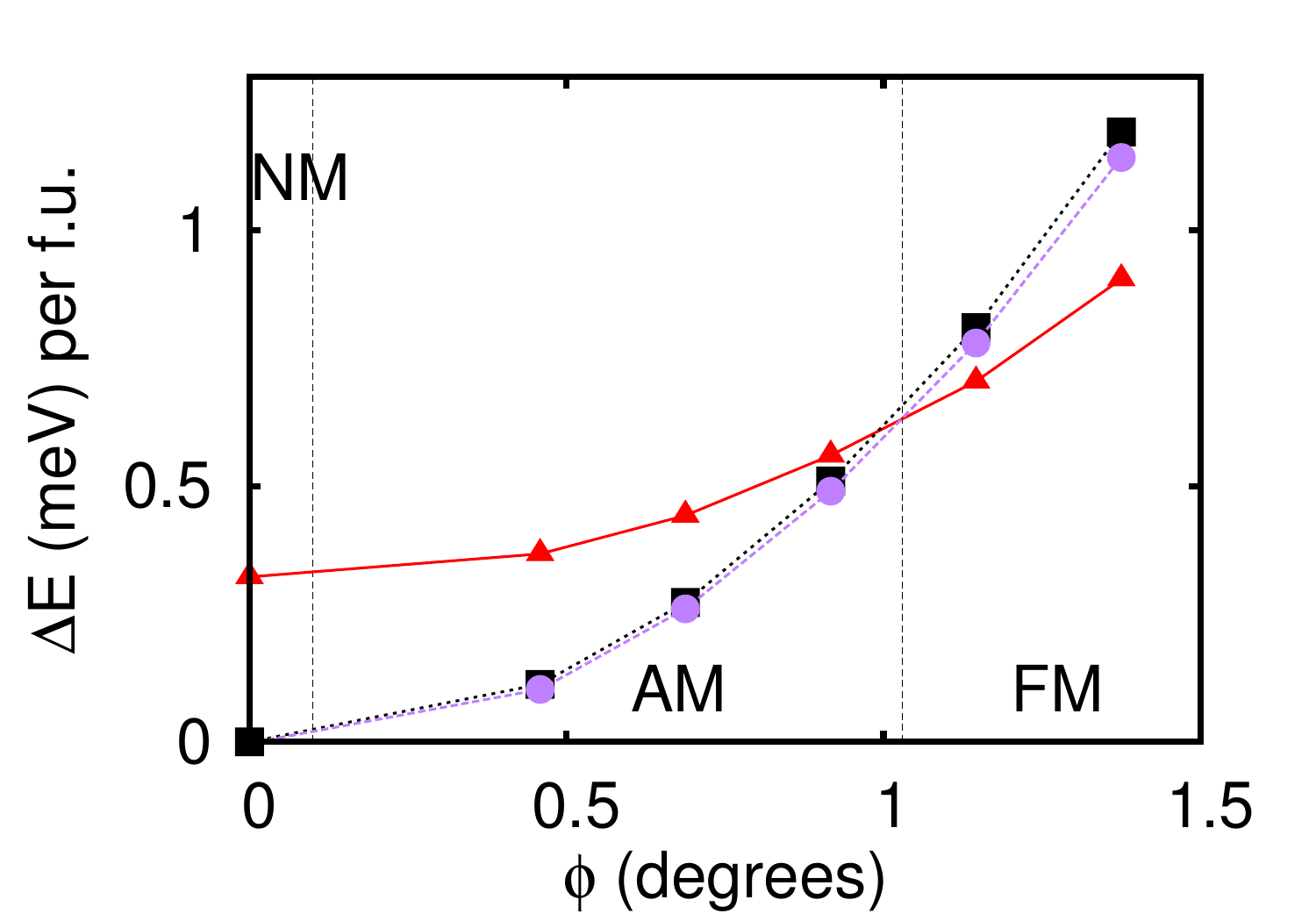}
\caption{Ground state energies as a function of the in-plane octahedral rotation angle $\phi$ for $U=0.52$~eV in the non-relativistic limit with no SOC with non-magnetic (NM, black squares), altermagnetic (AM, purple circles), and ferromagnetic (FM, red triangles) in phases.}\label{figure15}
\end{figure}

In the main text, we report the magnetic phase diagram in Fig.~\ref{figure6}. In this Appendix, we present an example of the underlying calculations by plotting in Fig.~\ref{figure15} the total energies for the non-magnetic (NM), altermagnet (AM), and ferromagnetic (FM) phases as a function of the in-plane octahedral angle $\phi$ for a fixed $U=0.52$~eV. Plots for other $U$ values in the regime where AM ordering is found present very similar features.
The energies of all three phases have a parabolic dependence on the in-plane rotation angle $\phi$, which follows the behavior of thermodynamic Landau theory, with $\phi$ as the order parameter. Notably, we find that the ground state changes even with small variations of $\phi$. Increasing $\phi$ even minutely from zero, the systems present an NM, AM, and finally an FM ground state. 

\section{Band structure plot in bulk Sr$_2$RuO$_4$ and estimation of spin-splitting}
\label{app:bulkband}
\begin{figure}[htb]
\centering
\includegraphics[width=6.2cm,height=\columnwidth,angle=270]{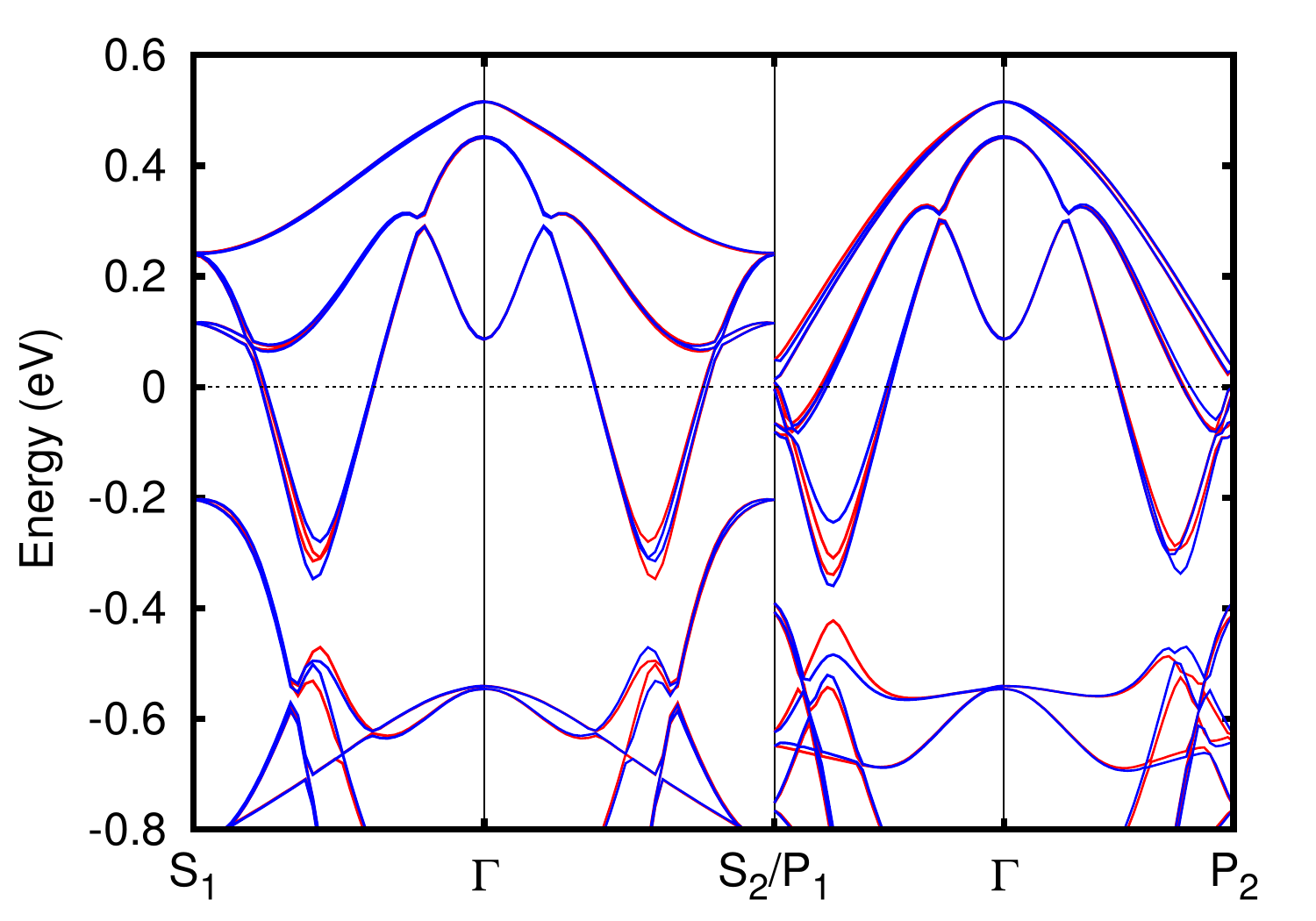}
\caption{Band structure of bulk Sr$_2$RuO$_4$ for $U=1.0$ eV and $\phi = 8^{\circ}$, with spin-up bands (blue) and spin-down bands (red) in the non-relativistic limit with no SOC. Similar to Fig.~\ref{BS_singlelayer} but here in the bulk.}\label{BS_bulk}
\end{figure}
In this Appendix, we provide in Fig.~\ref{BS_bulk} the overall band structure of altermagnetic bulk Sr$_2$RuO$_4$, along the same path in the Brillouin zone as for the single layer reported in Fig.~\ref{BS_singlelayer} in the main text. 
We use the same Coulomb repulsion $U=1.0$~eV and octahedral rotation angle $\phi=8^{\circ}$ to directly be comparable with the Fermi surface results in Fig~\ref{FS_bulk} in the main text. 
As discussed in the main text, the interlayer hybridization makes the points P$_1$ and P$_2$ asymmetric in the bulk due to the altermagnetic order and its non-relativistic spin-splitting. A finite spin-splitting also appears along $\Gamma$-S.
Without interlayer hybridization and magnetism, we would instead have fourfold degenerate bands due to Kramers' and layer degeneracies. 
Still, with relatively small interlayer hybridization, we expect the quartet of bands to be close to each other but separated by small splittings: one splitting of magnetic origin and another due to interlayer hybridization. As a consequence, the average of the spin-up eigenvalues is still close to the average of spin-down eigenvalues along the $\Gamma-\rm{S}$ path, while $\Gamma-\rm{P}$ shows more substantial spin-splitting.

\begin{figure}[htb]
\centering
\includegraphics[width=6.2cm,height=\columnwidth,angle=270]{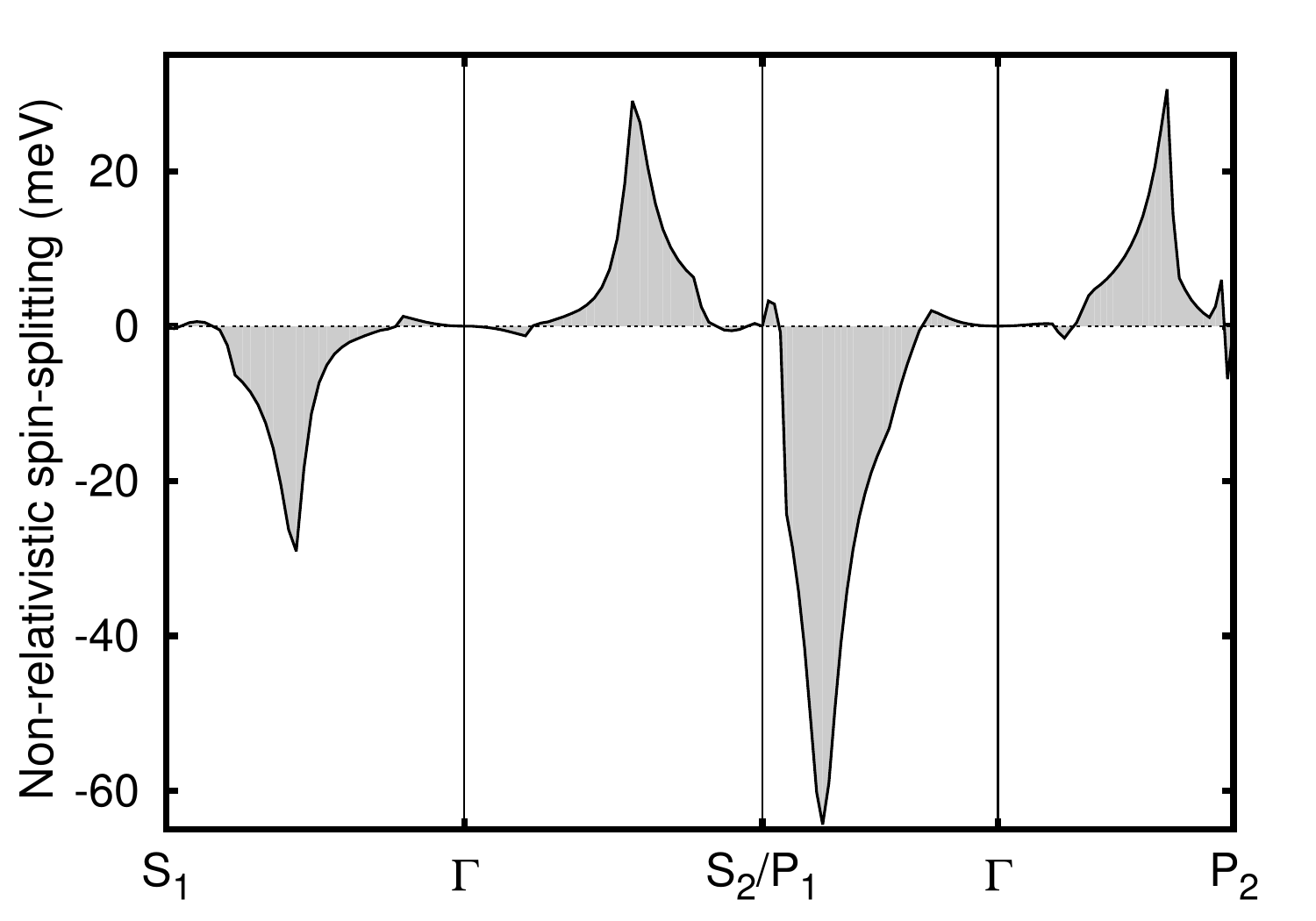}
\caption{Spin-splitting in bulk Sr$_2$RuO$_4$ for $U=1.0$~eV and $\phi=8^{\circ}$ in the non-relativistic limit with no SOC for the top bands of the quartet of bands crossing the Fermi level in Fig.~\ref{BS_bulk}.}
\label{NRSS_bulk}
\end{figure}

\begin{figure}[htb]
\centering
\includegraphics[width=6.2cm,height=\columnwidth,angle=270]{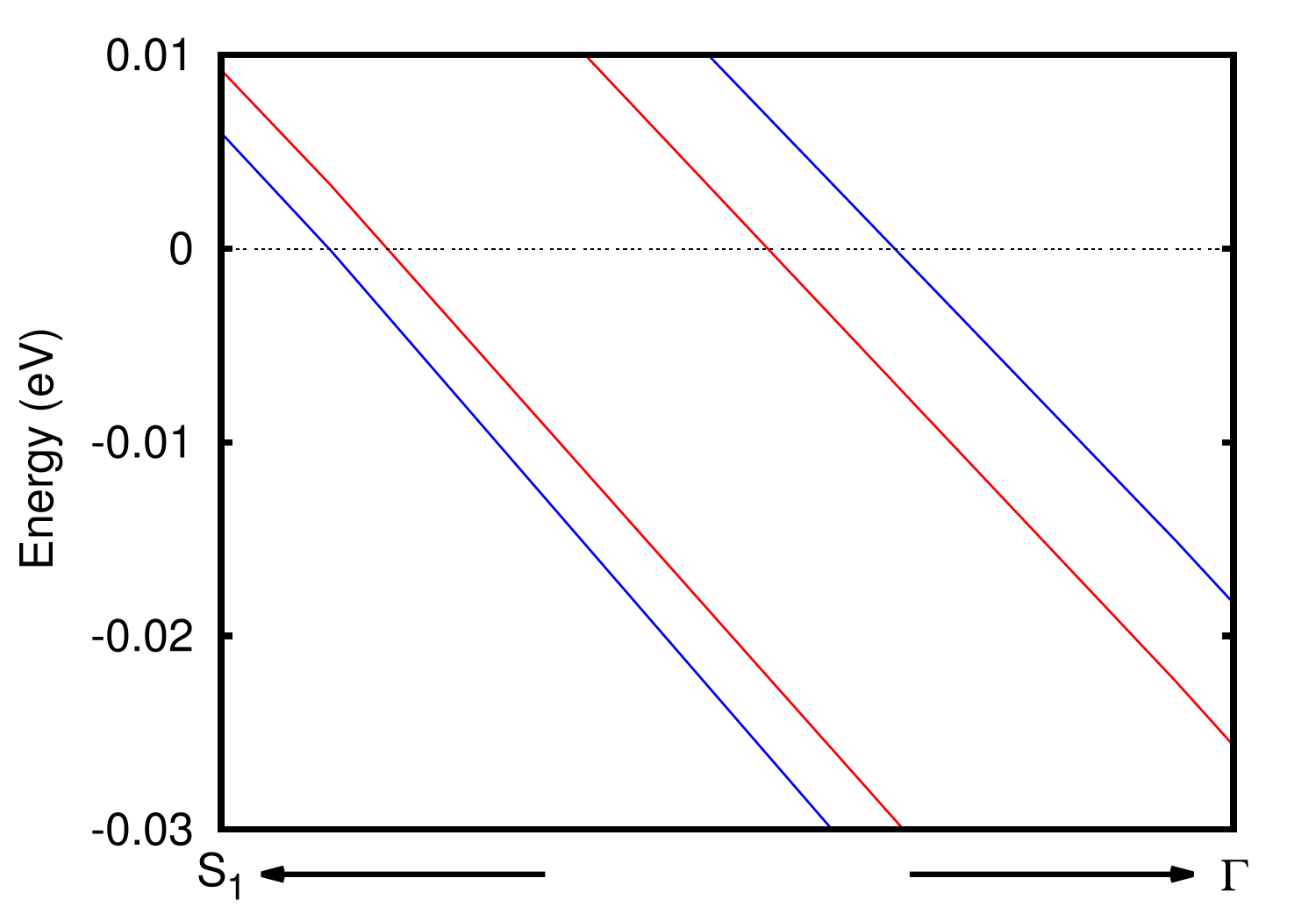}
\caption{Band structure in bulk Sr$_2$RuO$_4$ for $U=1.0$~eV and $\phi$ = 8$^{\circ}$ at the Fermi level (zero energy, black dashed line) along the path $\Gamma-\rm{S}_1$, with spin-up bands (blue) and spin-down bands (red) in the non-relativistic limit with no SOC. The extracted spin-splitting is approximately $3$ and $7$~meV for the $\alpha$ and $\beta$ Fermi surfaces, respectively.
}\label{NRSS_Fermi}
\end{figure}

To quantify the non-relativistic spin-splitting, we extract it along the same path as the band structure in Fig.~\ref{BS_bulk} and plot the results in Fig.~\ref{NRSS_bulk} for the spin-splitting of the two upper bands of the quartet crossing the Fermi level. 
As seen, the spin-splitting is larger along $\Gamma-\rm{P}$, but also clearly present along $\Gamma-\rm{S}$. Notably, along all the $k$-paths, the non-relativistic spin-splitting reaches a maximum approximately halfway between the $\Gamma$ point and the edge of the Brillouin zone, as is common in altermagnets.

Finally, in Fig.~\ref{NRSS_Fermi} we also extract the spin-splitting at the Fermi level, as that is important to understand the influence of the altermagnetic state on superconductivity. We find a non-relativistic altermagnet spin-splitting of $3$ and $7$~meV for the two bands crossing the Fermi level along $\Gamma-\rm{S}$, as displayed in Fig.~\ref{NRSS_Fermi}. These are thus the spin-splitting on the $\alpha$ and $\beta$ Fermi surfaces, respectively.
Similar spin-splitting occurs along $\Gamma-\rm{P}$. 
While we are here exaggerating the in-plane rotation angle $\phi={8}^{\circ}$, as in most of this work, in order to make the spin-splitting easier to visualize, we note that with the spin-splitting proportional to $\sin(\phi) \sim \phi$, a sizable spin-splitting will even be occurring for smaller rotation angles. In fact, as long as the spin-splitting is comparable or within the range of the superconducting gap estimated to be 350~$\mu$eV, the altermagnetic state will have consequences for superconductivity. 

\newpage
\bibliography{altermagnetism20nov}
\end{document}